\documentclass{appolbnh}

\usepackage{amsmath}
\usepackage{slashed}
\usepackage{multirow}
\usepackage{graphicx}

\begin{document}
\title{Odyssey of the elusive $\Theta^+$\footnote{Submitted to Acta Physica Polonica B special volume dedicated to
Dmitry Diakonov, Victor Petrov and Maxim Polyakov.}}
\author{Micha{\l} Prasza{\l}owicz
\address{Institute of Theoretical Physics\\ Faculty of Physics, Astronomy and Applied Computer Science\\ Jagiellonian University, \\
ul. S. {\L}ojasiewicza 11, 30-348 Krak{\'o}w, Poland.}}

\maketitle
\begin{abstract}
$\Theta^+$ is a putative light pentaquark state of positive parity with minimal quark content $(uudd\bar{s})$. It
naturally emerges in  chiral models for baryons, but experimental evidence is uncertain. We review the theoretical
foundations of  chiral models and their phenomenological applications to  exotic states.
In particular, we discuss in detail the pentaquark widths with special emphasis on the cancellations occurring
in the decay operator. We also discuss some experiments, mainly those whose positive evidence of 
${\mit\Theta}^+$ persists to this day.
This review is dedicated to Dmitry Diakonov, Victor Petrov, and Maxim Polyakov and their contribution to the
${\mit\Theta}^+$ story.
\end{abstract}

\section{Prologue}

In February 1987, I co-organized the {\em Workshop on Skyrmions and Ano{\-}ma{\-}lies}, 
which was held in a small palace in Mogilany near Kraków. There were 63 participants, among them Mitya Diakonov, whom I knew from his groundbreaking work on perturbative QCD (the so-called DDT paper~\cite{DDT}), but whom I did not have the opportunity to meet in person. 
Organizing an international conference in Poland at that time was an unusual challenge. The country was still recovering from martial law, and political pressure and constant suspicion of the authorities accompanied us at every stage of the workshop preparations. To this day, I do not know how it was possible that participants from Israel and South Korea,
countries with which communist Poland had no diplomatic relations, were granted Polish 
visas, and citizen Diakonov was allowed to travel to rebellious Poland.

Entirely immersed in administrative work, I completely forgot that I also had to prepare a paper. More than two years earlier,
we published an article on  SU(3) Skyrmion~\cite{Mazur:1984yf}, but by community standards, it was an
old result and I should have prepared something new. Browsing the literature, I came across predictions of
an exotic pentaquark belonging to the SU(3) flavor antidecuplet. I realized that it was  possible to constrain
its mass in a model-independent way based on non-exotic baryon data alone if one applied the second-order perturbation theory in the chiral symmetry breaker. The result was surprisingly low~\cite{Praszalowicz:1987em}: approximately 1535~MeV! 
At the time, I believed that the antidecuplet was beyond the model's area of applicability, especially since it seemed that the decay width should be quite large, of the order of 100--200~MeV. This gradually changed after conversations with Mitya, who invited me to Leningrad to discuss the SU(3) quantization of the chiral quark-soliton model that he and his colleagues derived from the instanton model of the QCD vacuum.

I went to Leningrad very quickly, meeting Vitya Petrov and also Pasha Pobylitsa. While discussing SU(3) quantization
of the chiral quark-soliton model, I discovered that it had a much reacher structure than the Skyrmion, and that some terms,
non-leading in the large-$N_c$ expansion, arise in this model naturally, while they had to be added {\em by hand}
in the Skyrme model~\cite{Guadagnini:1983uv}. Years later, when such terms were calculated for the decay operator,
it was found that they make the width of the pentaquark close to zero~\cite{Diakonov:1997mm}! However, at the moment, SU(3)
exotica were put on hold for 10 years and we worked with Mitya and Vitya (who visited me in Kraków 
in spring 1988) on other aspects of the chiral quark-soliton model.

In 1997, I was visiting Bochum University and met Maxim Polyakov, a young student of Mitya and Vitya at the time, who
was assigned to recalculate the pentaquark. I looked at the project with sympathy, but could not get over my 
skepticism that the model might not apply to the higher representations of the SU(3) flavor group. When they found
that the width was small, I immediately converted to the old ``religion''.  At the time, I did not even realize how seriously 
Mitya took this result, urging experimentalists to confirm it experimentally. It must have been early spring 2003 when I visited Bochum again, and Maxim dropped by my office to explain how Fermi motion was estimated in Nakano's paper~\cite{LEPS:2003wug}
describing the observation of~${\mit\Theta}^+$. ``Mission accomplished'' --- we thought.

Over the years we have become friends with Mitya, Vitya, and Maxim, both professionally and privately. In this review,
which is an expanded version of the Corfu 2023 proceedings~\cite{Praszalowicz:2024zsy}, 
I try to bring back the story of~${\mit\Theta}^+$, 21 years
after its first experimental announcements.

\newpage

\section{Introduction}

\begin{figure}[htb]
\centerline{%
\includegraphics[width=8.8cm]{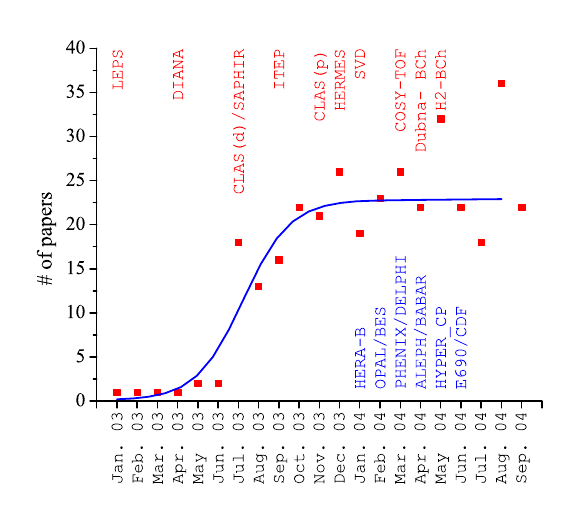}}
\caption{Number of pentaquark papers in \texttt{arXiv} per month in the first 21 months after the
publication of LEPS and DIANA. At the top in red, experimental papers confirming
pentaquark discovery, at the bottom in blue, no-observation experimental papers.
The blue solid line is for eye-guiding.}
\label{fig:papers}
\end{figure}

In 2003, two experimental groups, LEPS~\cite{LEPS:2003wug}  and DIANA~\cite{DIANA:2003uet}, announced the discovery
of a light, narrow, exotic baryon with a mass within the range of $1540$~MeV, 
which was later dubbed as ${\mit\Theta}^+$. Both groups concluded that the observed state may have been 
the lightest member of the antidecuplet of exotic pentaquark baryons, namely a $ u u d d \bar{s}$ state, which naively may be viewed as a $K$-nucleon system.
These experimental searches have been motivated by chiral models, which almost two decades earlier predicted light pentaquark $\overline{\bf 10}$ flavor
multiplet of positive parity.
The LEPS paper was submitted 
to \texttt{arXiv} on January 14, but the results were presented earlier at the PANIC Conference in Osaka in September/October 2002,
and the DIANA results were presented at the Session of Nuclear Division of the Russian Academy of Sciences  on
December 3, 2002. Nevertheless, little attention has been paid to these papers by the particle physics community until July 2003. On July 1, 
{\em The New York Times}~\cite{NYT2003}, and to the author's best knowledge, {\em USA Today}, published articles on the pentaquark discovery. This is
illustrated in Fig.~\ref{fig:papers}, where we plot the number  of pentaquark papers in \texttt{arXiv} per month from January 2003 until September 2004.

\vspace{-2mm}

We see that before July 2003 basically there was only one paper per month submitted to \texttt{arXiv}, among them
the diquark--triquark model by Karliner and Lipkin~\cite{Karliner:2003dt} submitted in February, the paper on photoexcitation
of antidecuplet by Polyakov and Rathke~\cite{Polyakov:2003dx} (March), or the paper from April by Walliser and Kopeliovich~\cite{Walliser:2003dy}
on exotica in topological soliton models. For obvious reasons, most of the papers after the July explosion 
were theoretical, although very soon
experimental analyses were quickly, sometimes too quickly, completed and submitted to \texttt{arXiv}.~In Fig.~\ref{fig:papers} we display
positive experimental papers in red (top) and the negative ones in blue (bottom). 

Obviously, none of these experiments, including LEPS and DIANA, were designed to search for pentaquarks. People used data
collected for other purposes. Only later were dedicated experiments  conducted with, 
however, mixed results. In 2004, ${\mit\Theta}^+$
paved its way to the Particle Data Group (PDG) listings~\cite{PDG2004} as a three-star resonance, 
in 2005 its significance was reduced to two stars,
and in 2007, it was omitted from the summary tables.~As of 2008, it is no longer listed by the PDG~\cite{PDG2008}.

Clearly, most non-observation experiments do not really exclude the existence of ${\mit\Theta}^+$, but rather put an upper limit on its production cross-section.
The cleanest and decisive experiment would be the so-called  {\em formation experiment} where the resonance is directly produced in the $K+N$ reaction.
DIANA is exactly this kind of experiment where the liquid xenon bubble chamber was exposed to a separated $K^+$ beam. On the contrary, LEPS
was a photoproduction experiment on the $^{12}$C carbon nucleus. 
In the follow-up analyses, both DIANA~\cite{DIANA:2006ypd,DIANA:2009rzq,DIANA:2013mhv} and 
LEPS in the dedicated photoproduction experiment on deuteron~\cite{LEPS:2008ghm,Nakano:2010zz} confirmed their initial findings.
We describe both experiments in more detail in Section~\ref{sec:exps}. An interesting analysis of why ${\mit\Theta}^+$ could be observed in some experiments and not in others can be found in Ref.~\cite{Azimov:2007hq}.

The experimental searches were inspired mostly by Mitya Diakonov's long-term efforts~\cite{VityaYouTube} to convince various experimental groups 
to risk time and reputation in search of the elusive pentaquark. Diakonov together with Vitya Petrov and Maxim Polyakov
co-authored a seminal paper on the mass and width of ${\mit\Theta}^+$ in the Chiral Quark-Soliton Model ($\chi$QSM)~\cite{Diakonov:1997mm} 
that approaches 1000 citations in 
{\tt InSpire.hep}\footnote{All three authors of this work published in \textit{Zeitschrift f{\"ur} Physik A} died prematurely: Diakonov in 2012
at the age of 63, Polyakov and Petrov in 2021 at the age of 55 and 66 respectively. \textit{Zeitschrift f{\"ur} Physik} does not exist anymore
as a separate journal. In 1997, it became a part of the  \textit{European Physical Journal}.}. The fact that exotic pentaquarks are generically light
in chiral models had been known already since the eighties, however the small width reported in~\cite{Diakonov:1997mm} was a real breakthrough.
In fact, the estimate of 15~MeV turned out to be too generous (today we know that the width must be smaller than 0.5 MeV), however, it was within
the accuracy range of the first experimental reports.

In the present paper, we want to recall the main theoretical and experimental facts about ${\mit\Theta}^+$
updating the analysis of Ref.~\cite{Ellis:2004uz}  published 20 years ago in 2004 and expanding a recent
review from 2023~\cite{Praszalowicz:2024zsy}.
This author firmly believes that ${\mit\Theta}^+$ story 
is not closed and that the pendulum of history may soon swing to the other side. And if so, these few comments may be useful for
those who are too young to remember how it all happened.

\section{Quark model}

Already Gell-Mann at the dawn of the quark model pointed out the possibility of exotica: pentaquarks  as well tetraquarks~\cite{Gell-Mann:1962yej,Gell-Mann:1964ewy}.
Of course, no dynamical calculations or phenomenological estimates of pentaquark masses were performed at the time. One can, however, relatively
easily perform such an estimate. Assuming that the constituent light quark mass is 1/3 of the nucleon mass, \ie\  approximately $M_q=313~$MeV,
and the constituent strange quark mass is 1/3 of the ${\mit\Omega}^-$ mass; \ie\  approximately $M_s=557~$MeV, we arrive at a rough estimate of the
lightest pentaquark state $u u d d \bar{s}$ of 1800~MeV. Alternatively, one could estimate the strange quark mass from the ${\mit\Xi}$--nucleon mass difference,
which is equal to 380~MeV obtaining $M_s=503~$MeV, leading to ${\mit\Theta}^+$ mass of 1755~MeV. Such states would, however, have negative parity $P$,
while chiral models predict $P=+$.
More sophisticated models of multiquark states were discussed within the framework of the bag model already in the late 
seventies~\cite{Jaffe_exotica,Strottman:1979qu}. However, to the best of our knowledge, no antidecuplet positive parity states 
were considered
at the time.

A few searches of strangeness $S=+1$ baryonic 
resonances have
been carried out in the above mass range with the null result reported in the 1986 edition of the PDG listings~\cite{ParticleDataGroup:1986kuw} with the following
comment: {\em The evidence for strangeness +1 baryon resonance was reviewed in our 1976 edition (...). The general prejudice against baryons not made of
three quarks and the lack of any experimental activity in this area make it likely that it will be another 15 years before the issue is decided.} 

${\mit\Theta}^+$ decay modes are $K^0 p$ or $K^+ n$. For the masses given above, the kaon momentum in the ${\mit\Theta}^+$ rest frame
is within the range of $p= 490 \div 530~$MeV. If ${\mit\Theta}^+$ has the negative parity, we can take as a benchmark $N(1535)$ nucleon resonance of
spin 1/2 and total width $\sim 150~$MeV. Since the pion momentum in the decay of $N(1535)$ to $\pi N$ is approximately 460 MeV, and the decay is
in $s$-wave, we naively expect the ${\mit\Theta}^+$ decay width to be of the same order, approximately 10\% larger. 
If the parity of ${\mit\Theta}^+$ is positive, we can use ${\mit\Delta}$~resonance to estimate 
its width. ${\mit\Delta}$ width is approximately 120~MeV, and the pion momentum in the $p$-wave decay ${\mit\Delta} \rightarrow \pi N$ is $p=227~$MeV. In this case,
the width scales as a third power of $p$, so we expect the pentaquark width to be 10 times larger than the one of ${\mit\Delta}$! In any case, the naive
quark model predicts heavy and wide exotica, which have not been confirmed experimentally~\cite{ParticleDataGroup:1986kuw}.

The minimal quark content of ${\mit\Theta} ^{+}$  is $
uudd\bar{s} $.
Two quarks can be either in flavor $%
\overline{\bf 3}$ or ${\bf 6}$ 
\begin{equation}
{\bf 3}\otimes {\bf  3} =\overline{\bf  3}\oplus {\bf  6} \,.
\end{equation}
Therefore, possible representations for 4 quarks are
contained in the direct product%
\begin{equation}
    \left(\,\overline{\bf  3}\oplus {\bf  6}\right)\otimes\left(\,\overline{\bf  3}\oplus {\bf  6}\right)\rightarrow {\bf 3}\oplus \overline{\bf 6}
   \oplus{\bf15}\oplus{\bf 15}^{\prime }\,.  \label{twodiqs}
\end{equation}%
Here, ${\bf15}=(2,1)$ and ${\bf 15}^{\prime }=(4,0)$\,.
Adding a $\overline{\bf 3}$ antiquark
yields%
\begin{align}
{\bf 3}\otimes \overline{\bf 3}& ={\bf 1}\oplus{\bf 8}\,,  \nonumber \\
\overline{\bf 6}\otimes \overline{\bf 3}& ={\bf 8}\oplus\overline{\bf 10}\,,  \nonumber \\
    {\bf 15} \otimes \overline{\bf 3}& ={ \bf 8}\oplus{\bf 10}\oplus{\bf 27}\,,  \nonumber \\
{\bf 15}^{\prime }\otimes \overline{\bf 3}& ={\bf 10}\oplus{\bf 35}\,,  \label{last}
\end{align}%
Therefore, the $\left| q^{4}\overline{q}\right\rangle $ state can be in one of
the following flavor representations:
\begin{equation}
\left| q^{4}\overline{q}\right\rangle \in {\bf 1},\,{\bf 8},\,{\bf 10},\,\overline {\bf 10}%
,\,{\bf 27},\,{\bf 35}\,.   \label{reps}
\end{equation}
Whether all representations (\ref{reps}) are allowed  depends
on the dynamics of a specific model and on the constraints coming from the Pauli principle.

\begin{figure}[htb]
\centerline{%
\includegraphics[width=7.5cm]{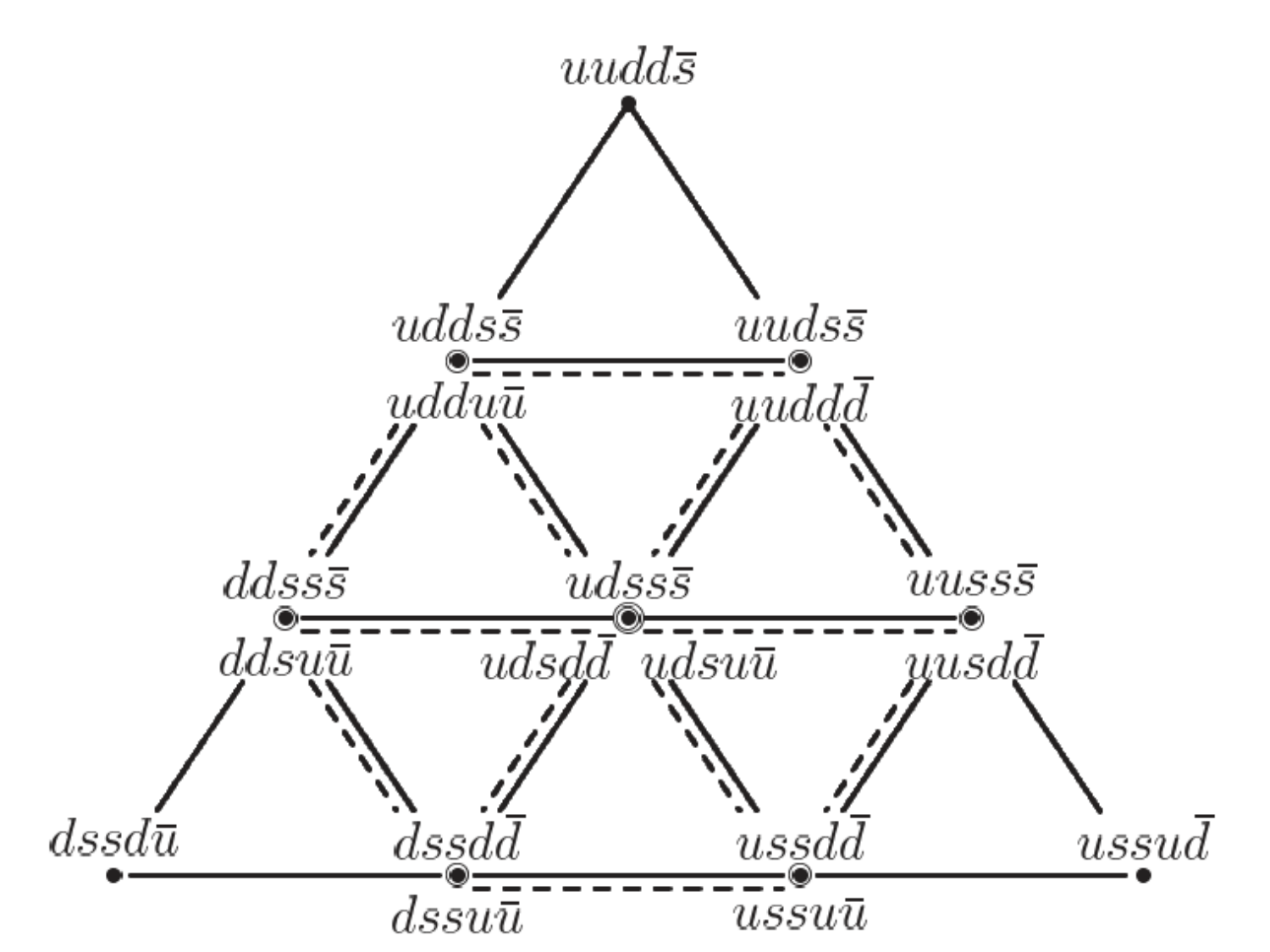}}\vspace{-1.5mm}
\caption{Pentaquark multiplets $\overline{\bf 10}$ (solid triangle) and ${\bf 8}$ (dashed octagon) that follow from the quark model.}%
\label{fig:mixing}
\end{figure}

Out of allowed representations (\ref{reps}), the lowest one including
explicitly exotic states is $\overline{\bf 10}$ which appears in a direct
product of 4 quarks in flavor $\overline{\bf 6}$ and an antiquark (\ref{last})
\begin{equation*}
\overline{\bf 6}\otimes\overline{\bf 3}\rightarrow {\bf 8}\oplus\overline{\bf 10}   \label{8and10bar}
\end{equation*}
and is, therefore, inevitably accompanied by an octet (see Fig.~\ref%
{fig:mixing}). Unlike in the case of the ordinary octet and
decuplet, pentaquark symmetry states (\ie\ states which are pure
octet or antidecuplet) do not have a unique quark structure~\cite{Praszalowicz:2004xh}. For example, a
proton-like state in antidecuplet and octet have the following quark content:%
\begin{align}
\left| p_{\overline{\bf 10}}\, \right\rangle & =\sqrt{\frac{2}{3}}\left| uuds\bar{s}\,\right\rangle +%
\sqrt{\frac{1}{3}}\left| uudd \bar{d}\,\right\rangle\,,  \nonumber \\
\left| p_{\bf 8}\, \right\rangle & =\sqrt{\frac{1}{3}}\left| uuds\bar{s}\,\right\rangle -\sqrt{\frac{%
2}{3}}\left| uudd \bar{d}\,\right\rangle\,,   \label{ideal}
\end{align}
where it is implicitly assumed that four quarks are in a pure $\overline{\bf 6}$
state. 
Similarly ${\mit\Sigma}$-like pentaquarks, ${\mit\Xi}^0$ and ${\mit\Xi}^-$ are mixtures of the pure quark states
analogous to (\ref{ideal}), while ${\mit\Theta}^{+}$ and ${\mit\Xi}^+$
and ${\mit\Xi}^{--}$
are the pure quark states (they correspond to three vertices of the $\overline{\bf 10}$ triangle --- see Fig.~\ref{fig:mixing}).
The latter ones are truly {\em exotic}, because their quantum numbers cannot be obtained from
three quarks only. The remaining
states in Fig.~\ref{fig:mixing}, although consisting also of  five 
quarks\footnote{Throughout this paper, we shall use the term {\em quark} both for quarks and antiquarks, unless
we explicitly need to distinguish the two.},
are {\em cryptoexotic} because their
quantum numbers can be obtained from three quarks. Therefore, they can mix with regular baryons.

\vspace{-2mm}

This observation led Jaffe and Wilczek~\cite{Jaffe:2003sg} to propose a diquark model for positive parity pentaquarks,
where the physical states would
correspond to pure quark states. This scenario was dubbed as an {\em ideal mixing}. Group theoretical considerations provide us with mass formulas
with a number of free parameters that have to be fixed from the data (see \eg\ \cite{Praszalowicz:2004xh}). Jaffe and Wilczek used obviously
the reported mass of ${\mit\Theta}^+$, and
two nucleon resonances, Roper $N(1440)$ and $N(1710)$, which
were associated with $\left| uudd \bar{q}\right\rangle$ and $\left| quds\bar{s}\right\rangle$ states, respectively, where $q=u$ or $d$. The problem with
this assignment was, however, that Roper and $N(1710)$ have very different partial widths to $\pi N$ 
($\sim230$ and 15~MeV, respectively~\cite{ParticleDataGroup:2024cfk}), while the ideal mixing scenario predicts that these widths are 
nearly the same~\cite{Cohen:2004gu,Praszalowicz:2004xh}.

If all quarks in $uudd\bar{s}$ were in the ground state, the parity of ${\mit\Theta}^+$ would be negative. However, soliton models 
that prompted experimental searches, predicted pentaquark parity to be positive. In the model of Jaffe and Wilczek, the quarks were strongly
correlated forming a spin-zero, and color and flavor $\overline{\bf 3}$ diquarks: $[ud]$, $[ds]$, and $[us]$. 
In order to form a color singlet with an antiquark, two diquarks have 
to be  antisymmetric in color $\overline{\bf 3} \otimes \overline{\bf 3}$ (\ie\ in a triplet), 
symmetric  in flavor (\ie\ in $\overline{\bf 6}$ as mentioned above),
and therefore antisymmetric in space, \ie\ in the negative space-parity configuration. When combined with an antiquark, the resulting
pentaquark has, therefore, positive~parity.

To circumvent the parity problem, Karliner and Lipkin~\cite{Karliner:2003dt} proposed a model with a {triquark}\footnote{In the case
of ${\mit\Theta}^+$, the ($ud\bar{s}$).} and a diquark correlations. In their model, the two clusters, a $[ud\,]$ diquark and a $(ud\bar{s})$ triquark were in a relative
$p$-wave. They argued that the $s$-wave configuration was suppressed due to the hyperfine repulsion between the two clusters.
In order to estimate the pentaquark masses, they used the Zeldovich--Sakharov model~\cite{Zeldovich}, where quarks interact 
through a color-magnetic force,
and various phenomenological inputs both from meson and baryon spectroscopy. Their mass estimate of ${\mit\Theta}^+$ was in rough agreement with LEPS and DIANA.

These quark pentaquark models were proposed after the announcement of ${\mit\Theta}^+$. However, light, positive parity flavor $\overline{\bf 10}$
exotic multiplet was predicted much earlier within the framework of chiral models, which we discuss in the next sections.

\section{Chiral effective models for QCD}
\label{sec:effQCD}

In this section, we introduce soliton models for baryons. We first discuss the classical solution and identify its symmetries. Next, we show
how the soliton is quantized and which SU(3)$_{\rm flavor}$ representations emerge. We  describe how mass formulas and
decay widths can be computed. The Reader more interested in numerical predictions can skip ahead to Section~\ref{sec:pheno}.

Although there exists a fundamental theory of strong interactions, namely Quantum Chromodynamics (QCD), its practical applicability to the
low energy physics of hadrons is rather limited. One has to resort to computer simulations, which are technically difficult,
especially for five-quark operators and states above meson--baryon thresholds.
Indeed, early lattice QCD computations for ${\mit\Theta}^+$ were rather inconclusive~\cite{Chiu:2005is,Ishii:2004ib,Takahashi:2004sc,Sasaki:2004cg}. 
Therefore, instead of solving QCD, one constructs effective models that share the symmetries of QCD  and are technically tractable.

The idea behind the effective models is to approximate the QCD Lagrangian for light quarks
\begin{equation}
{\cal L}_{\rm QCD}=\bar{\psi}\left( i \slashed{\partial} +g \slashed{\boldsymbol{A}}-m \right)\psi 
-\frac{1}{2}{\rm Tr} \boldsymbol{F}_{\mu \nu} \boldsymbol{F}^{\mu \nu}
\label{eq:LQCD}
\end{equation}
in terms of different degrees of freedom and different interactions. Here, 
$\psi_{\alpha}=(u_{\alpha},d_{\alpha},s_{\alpha})$ is a flavor SU(3) vector constructed from the light quark Dirac bispinors of color ${\alpha}=1,2,\ldots,N_c$,
$\boldsymbol{A}_{\mu}=T^a A^a_{\mu}$ denotes an octet of gluon fields, and 
$\boldsymbol{F}_{\mu \nu}=T^a F^a_{\mu \nu}$ is the QCD field tensor. $T^a$ stand for the color
SU(3) generators. The quark mass matrix $m={\rm diag}(m_u,m_d,m_s)$ is considered to be a small perturbation and is
 set to zero in the chiral limit.

In the chiral limit, left- and right-handed quarks transform independently under global ${\rm SU}_{\rm L,R}(3)$ transformations, and it is well known that this symmetry
is broken to the vector subgroup ${\rm SU}_{\rm R+L}(3)$ by the vacuum state. The breakdown of chiral symmetry leads to the nonzero
quark condensate $\langle 0 |\bar{\psi}\psi | 0 \rangle$,
to the emergence
of Goldstone bosons, and to the dynamical generation of a constituent quark mass $M\sim 350$~MeV.

\subsection{Chiral quark model}

One can imagine that we integrate out  gluon fields from (\ref{eq:LQCD})  
and are, therefore, left with the quark degrees of freedom only. 
The quarks will still have canonical 
kinetic energy and possibly a mass term, 
however, interaction Lagrangian will consist of an infinite number of nonlocal many-quark vertices which, however, will
be chirally invariant. 
Typically, one  truncates this Lagrangian to the local four-quark interaction, the so-called 
Nambu--Jona-Lasinio model~\cite{Nambu:1961tp,Nambu:1961fr}. To ensure chiral invariance,
it is convenient to introduce eight auxiliary pseudo-Goldstone fields $\boldsymbol{\varphi}$ (pions, kaons, and $\eta$)
in a form of a unitary SU(3) matrix
\begin{equation}
U=\exp\left(i  \frac{2\boldsymbol{\lambda}\cdot \boldsymbol{\varphi}}{F} \right) \,,
\end{equation}
where $\boldsymbol{\lambda}$ are Gell-Mann matrices and $F$ is a pseudoscalar (pion) decay constant that in the present normalization
is equal to 186~MeV.

The simplest Lagrangian following from the above procedure, the chiral quark model Lagrangian,  is given by
\begin{equation}
{\cal L}_{\chi{\rm QM}}=\bar{\psi}\left( i \slashed{\partial}-m -M U^{\gamma_5} \right)\psi \,,
\label{chiQM}
\end{equation}
where
\begin{equation}
U^{\gamma_5}= U \frac{1+\gamma^5}{2}+ U^{\dagger}  \frac{1-\gamma^5}{2}\,.
\end{equation}

This remarkably simple Lagrangian was in fact derived~\cite{Diakonov:1985eg,DiakonovMogilany}
in the mid-eighties
from the instanton picture of the QCD vacuum~\cite{Diakonov:1983hh,Shuryak:1983ni}. Original
instanton-based calculations yield a momentum-dependent constituent quark mass $M(p)$ that
vanishes for large momenta and tends to $\sim 350$~MeV for $p=0$.

Saturating multiquark chiral interactions with Goldstone fields only is of course an approximation. A complete Lagrangian
including scalar, pseudoscalar, vector, axial, and tensor fields was constructed in Ref.~\cite{Diakonov:2013qta}, but
was never used in practical calculations.

A few comments concerning Lagrangian (\ref{chiQM}) are in order. ${\cal L}_{\chi{\rm QM}}$ is color diagonal, so it is formally proportional to
$N_c$ when summed up over  color indices $\alpha$. Chiral interactions given by the last term in  (\ref{chiQM}) do not confine, so this
very important feature of QCD has been lost. 
There is no kinetic part for the Goldstone bosons, which are merely  quark bilinears, as far as
the pertinent equations of motion are concerned. 

\vspace{-2mm}
\subsection{Skyrme model}

The kinetic part for the Goldstone bosons appears when we integrate out quarks 
\cite{Diakonov:1983bny,Balog:1984upv,Praszalowicz:1989dh,RuizArriola:1991gc} ending up with a Lagrangian given in terms of the Goldstone bosons
alone. This Lagrangian is organized as a power series in Goldstone boson momenta, \ie\  in terms of $\partial_{\mu}U$.
Such Lagragians are used for precision calculations in the chiral perturbation theory~\cite{Scherer:2002tk}.

The first term in $\partial_{\mu}U$ expansion, a quadratic term, is fully dictated by the chiral symmetry and is known as the Weinberg
Lagrangian~\cite{Weinberg:1978kz}. Higher-order terms of known group structure have, however, free coefficients that are not constrained
by any symmetry and have to be extracted from experimental data. Obviously, once we have at our disposal a reliable Lagrangian
like (\ref{chiQM}), we can compute the effective Goldstone boson Lagrangian to any order in $\partial_{\mu}U$.

In 1961, Skyrme~\cite{Skyrme:1961vq,Skyrme:1962vh} proposed the effective Goldstone boson Lagrangian that was later generalized by Witten~\cite{Witten:1983tw,Witten:1983tx}, which takes the following form:
\begin{equation}
{\cal L}_{\rm Sk}  = \frac{F^{2}}{16} 
{\rm Tr}\left(\partial_{\mu}U^{\dagger}\partial^{\mu}U\right) 
+  \frac{1}{32e^{2}} 
{\rm Tr}\left(\left[\partial_{\mu}U\,U^{\dagger},\partial_{\nu}U\,U^{\dagger}\right]^{2}\right) +{\cal L}_{\rm m} \,.
\label{eq:LSkyrme}
\end{equation}
The first term in (\ref{eq:LSkyrme}) is the Weinberg Lagrangian, the second one is called the {\em Skyrme term}. 
Parameter $e$ can be inferred from the pion scattering and is of the order of $e=4\div6$.
A possible $4^\mathrm{th}$-order term symmetric
in $\partial_{\mu}U$ derivatives, the so-called {\em non-Skyrme term}, has been also considered in the literature~\cite{Pham:1985cr}. Mass term
Lagrangian
\begin{equation}
{\cal L}_{\rm m}  = 
a \, {\rm Tr}\left(U+U^{\dagger}-2\right)
 +  b \,  {\rm Tr}\left(\left(U+U^{\dagger}\right)\lambda_{8}\right)
\label{eq:mlagr}
\end{equation}
takes care of the chiral symmetry breaking. Coefficients $a$ and $b$ are given as combinations of pseudoscalar meson
masses and can be found \eg\ in Ref.~\cite{Praszalowicz:1991aj}.

Note that each term in (\ref{eq:LSkyrme}) can be expanded in powers of $\boldsymbol{\varphi}$ generating {\em perturbative} Goldstone boson
interactions involving any even number of $\varphi_a$ fields. Therefore, at first glance, it appears that (\ref{eq:LSkyrme}) has nothing to do
with baryons. However (\ref{eq:LSkyrme}) admits {\em nonperturbative} solutions, known as solitons, that can be interpreted as baryons.
Similarly, soliton solutions also exist for the system described by  the chiral Lagrangian of Eq.~(\ref{chiQM}).

In the following, we will discuss how exotic baryons emerge in this framework. Before that, let us add that in the case of 
 SU(3) flavor
symmetry the chiral action corresponding to Skyrme's Lagrangian (\ref{eq:LSkyrme}) has to be supplemented by 
the Wess--Zumino--Witten term
${\mit\Gamma}_{\rm WZ}$~\cite{Callan:1969sn,Wess:1971yu}
\begin{equation}
S_{\rm Sk}= \int \mathrm{d}t \,{\cal L}_{\rm Sk} + {\mit\Gamma}_{\rm WZ} \, ,
\end{equation}
which is related to the chiral anomaly and does not affect equations of motion. $ {\mit\Gamma}_{\rm WZ}$ is related to the
topology of the $\boldsymbol{\varphi}$ field~\cite{Witten:1983tw}. It was shown~\cite{Diakonov:1983bny} that it follows from the imaginary part of the
action obtained by integrating out the quark fields in (\ref{chiQM}).
$ {\mit\Gamma}_{\rm WZ}$ cannot be written in terms of a local 
Lagrangian
density; instead, it is given as an integral over the 5-dimensional manifold
whose boundary is a 4-dimensional space-time
\begin{equation}
{\mit\Gamma}_{\rm WZ}=-i \frac{N_c}{240\,\pi^{2}}
\int \mathrm{d}^{5}r \, \epsilon^{\mu \nu \rho \sigma \tau}
{\rm Tr}\left(\partial_{\mu}U\,U^{\dagger} \:
         \partial_{\nu}U\,U^{\dagger} \:
	  \partial_{\rho}U\,U^{\dagger} \:
	   \partial_{\sigma}U\,U^{\dagger} \:
	    \partial_{\tau}U\,U^{\dagger} \right)\,.
\label{eq:WZ}
\end{equation}
In fact, the fifth, 
redundant coordinate, can be integrated out for the soliton configuration.

\vspace{-2mm}
\subsection{Hedgehog symmetry and solitons}
\label{ssec:hedgehog}

For massless free quarks ($m=0$ and $M=0$), left and right fermions can be independently rotated by global SU(3) transformations
\begin{equation}
\psi_{\rm L} \rightarrow L \psi_{\rm L}\,,\qquad \psi_{\rm R} \rightarrow R \psi_{\rm R} \,
\label{eq:ChiraLR}
\end{equation}
leaving (\ref{chiQM}) invariant. Here,
\begin{equation}
\psi _{\rm{L,R}}=\frac{1}{2}\left(1\mp \gamma ^{5}\right)\psi \, .
\end{equation}
Transformations (\ref{eq:ChiraLR}) leave the interaction term invariant ($M\not= 0$) if
\begin{equation}
U \rightarrow LUR^{\dagger} \,,
\end{equation}
which is  nothing else but a nonlinear realization of chiral symmetry~\cite{Scherer:2002tk}. Vacum state
corresponding to $U=1$ (or $\boldsymbol{\varphi}=0$) breaks this SU$_{\rm L}$(3)$\otimes$ SU$_{\rm R}$(3)
symmetry to vector SU(3)
\begin{equation}
L=R \,.
\end{equation}

Matrix $U$ is both time- and space-dependent, $U=U(t,\boldsymbol{r})$. For static configurations, $U(\boldsymbol{r})$
can be viewed as a mapping $R^3 \rightarrow\,$SU(3). However, if we require that at spacial infinity $U(\boldsymbol{r}\rightarrow \infty)=1$,
\ie\ that the system tends to the vacuum state, then all points at spacial infinity can be squeezed into one point, changing
the topology of $R^3$ to the one of a three-sphere $S^3$. Mappings of $S^3 \rightarrow\,$SU(3) are characterized by a winding
number, since SU(3) (or more precisely any SU(2) subgroup of SU(3), \eg\ isospin) has also a topology of a three-sphere. Indeed, any SU(2)
matrix can be parametrized~as
\begin{equation}
U_{\rm SU(2)}= a_0+ i \, \boldsymbol{a}\cdot \boldsymbol{\tau} \, ,
\end{equation}
where

\vspace{-6mm}
\begin{equation}
\sum_{i=0}^{3} a_i^2=1 \, .
\label{eq:SU2sphere}
\end{equation}
Equation (\ref{eq:SU2sphere}) is an equation for a three-dimensional sphere of radius one. The winding number (or the topological number) counts
how many times the spacial three-sphere is wrapped around the SU(2) sphere. Such mappings fall into distinct
topology classes and one cannot go from one class to another by a continuous deformation. The winding number of the 
$U$-mapping reads as follows:

\vspace{-3mm}
\begin{equation}
N_{\text{w}}=\frac{1}{24\pi ^{2}}\varepsilon ^{ijk}\int \mathrm{d}^{3}r\,{\rm Tr}%
\left[ \left( U^{\dagger }\partial _{i}U\right) \left( U^{\dagger }\partial
_{j}U\right) \left( U^{\dagger }\partial _{k}U\right) \right]\,,   \label{Nw}
\end{equation}%
and we can see a clear relation to the Wess--Zumino--Witten term (\ref{eq:WZ}).

The SU(2) mappings that have a nontrivial topological number can be represented in the form of a {\em hedgehog} Ansatz
\begin{equation}
u_0=\exp \left( i\,\boldsymbol{n}\cdot \boldsymbol{\tau}%
\,P(r)\right) \,,
\label{eq:hedg}
\end{equation}
where $\boldsymbol{n}=\boldsymbol{r}/r$. Function $P(r)$ has to vanish at infinity, so that $u_0 \rightarrow 1$. For the
hedgehog Ansatz (\ref{eq:hedg}) $N_{\rm w}=P(0)/\pi$. Thus, we conclude that for a soliton of $N_{\rm w}=1$, function $P$ has to satisfy
$P(0)=\pi$.
The hedgehog Ansatz (\ref{eq:hedg}) has a very special property: any spacial rotation of the unit vector $\boldsymbol{n}$ can be undone
by an internal SU(2) (isospin) rotation acting on Pauli matrices $\boldsymbol{\tau}$. 
This property is called {\em hedgehog symmetry}.
Exactly for this reason, mapping (\ref{eq:hedg}) has a nontrivial
winding number. 

In a seminal paper from 1979, Witten suggested that baryons may emerge as solitons in the
effective theory of mesons ~\cite{Witten:1979kh}, which in turn emerges from QCD in the large-$N_c$ limit. The simplest choice for such a theory is
the Skyrme Lagrangian (\ref{eq:LSkyrme}), where $U\!=\!u_0$ of Eq.~(\ref{eq:hedg}) and the winding number  $N_{\rm w}$ is 
interpreted as a baryon number. Euler--Lagrange equations of motion
reduce in this case to a differential equation for $P(r)$~\cite{Adkins:1983ya}.~As a result, we end up with the solitonic static solution
of mass $M_{\rm sol}\!\sim\! N_c$ (expressed in terms of space integrals over the function $P(r)$ with the energy determined by the meson decay constant $F$)
and baryon number equal to one, which we will call a {\em classical} baryon. The boundary condition $P(0)\!=\!\pi$ is required not only to ensure
that the winding number is equal to one, but it is also necessary for the soliton energy to be finite.
We will shortly explain how flavor and spin emerge in this picture.

In the chiral quark model (\ref{chiQM}), the soliton is a more complicated object. Here, in the limit of a large number  of colors
($N_{c}\! \rightarrow\!\infty$),  $N_{c}$ relativistic~va-\break lence quarks generate
chiral mean fields represented by a distortion of  the Dirac sea.~Such
distortion interacts with  valence quarks changing their wave function,
which in turn modifies the sea until a stable configuration is reached. This
configuration, called  \emph{chiral quark-soliton}, corresponds to the solution of the
Dirac equation following from (\ref{chiQM}) in the
mean-field approximation where the mean fields respect the
hedgehog symmetry, \ie\  with $U=u_0$.

Due to the hedgehog symmetry of $u_0$ neither  total angular momentum ($\boldsymbol{J}=\boldsymbol{L}+\boldsymbol{S}$) 
nor isospin ($\boldsymbol{T}$) are good symmetries
of Lagrangian (\ref{chiQM}). Instead, eigenvalues of grand spin $\boldsymbol{K}=\boldsymbol{J}+\boldsymbol{T}$ are good quantum numbers
for the soliton solution. The hedgehog quark state is intuitively described in Hosaka's article in this volume~\cite{Hosaka_DPP_volume}. 
The hedgehog symmetry emerges because it is impossible to construct a pseudoscalar field that changes 
a sign under inversion of coordinates,
which would be compatible with the $\mathrm{SU}(3)_{\rm flav}\otimes \mathrm{SO}(3)$ space symmetry. A smaller
hedgehog symmetry leads, as we shall see, to the correct baryon spectrum.
Since the valence level has $K=0$, the soliton solution carries no quantum numbers, except the
baryon number of the valence quarks. The number of valence levels depends on the topological number of the
mean field $u_0$, however topological condition $P(0)=\pi$ is not necessary for the soliton energy to be finite~\cite{Diakonov:1988mg}. 

\begin{figure}[htb]
\centerline{%
\includegraphics[width=7.6cm]{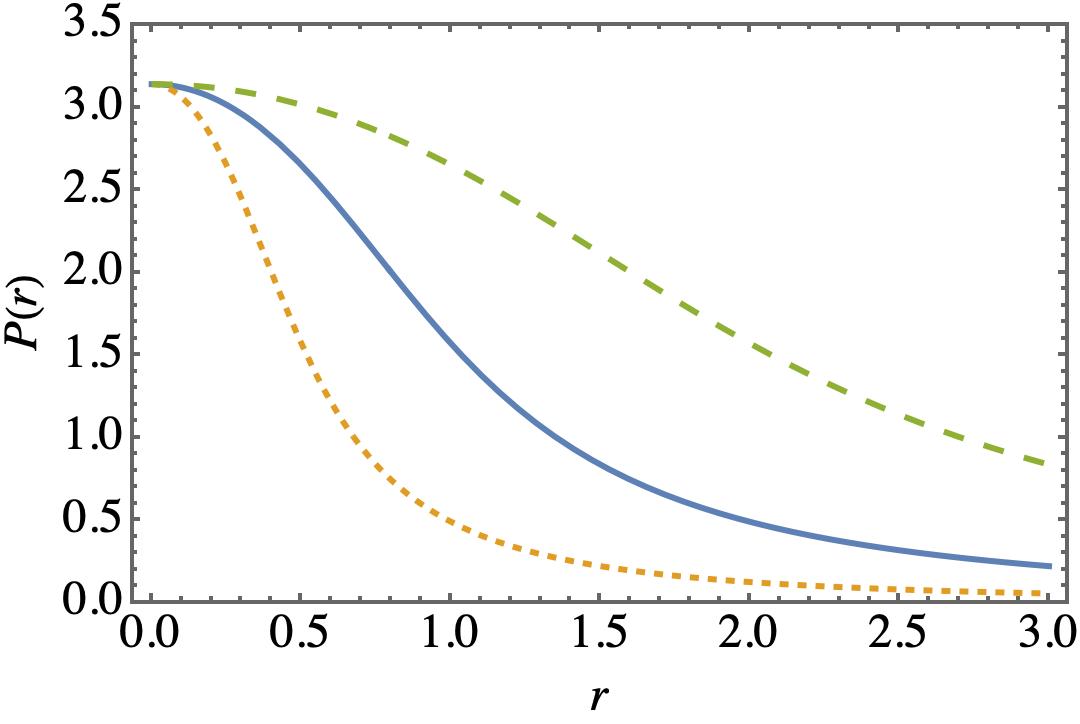}}\vspace{-2mm}
\caption{Soliton profile function $P(r)$  for $r_0=1/2$ (short-dashed orange),
$r_0=1$ (solid blue), $r_0=2$ (long-dashed green) in arbitrary units.}\vspace{-2mm}
\label{fig:profile}
\end{figure}

The best way to illustrate what happens (and in fact also for practical calculations) is to use the variation principle. To this
end, one uses an Ansatz for the profile function $P(r)$~\cite{Diakonov:1988mg}
\begin{equation}
P(r)=2\arctan\left[  \left(  \frac{r_{0}}{r}\right)  ^{2}\right]\,. \label{Pr}%
\end{equation}
This function is equal to $\pi$ at $r=0$ and vanishes at $r\rightarrow\infty$
as $r^{-2}$, which is the asymptotics  following from the pertinent Dirac equation. 
The variational parameter $r_0$ is called the soliton size. Function $P(r/r_0)$ is plotted in Fig.~\ref{fig:profile}.

\begin{figure}[htb]
\centerline{%
\includegraphics[width=9.4cm]{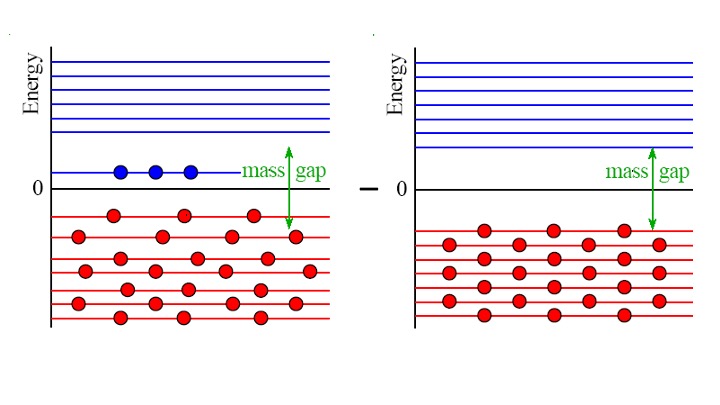}}
\centerline{%
\includegraphics[width=9.4cm]{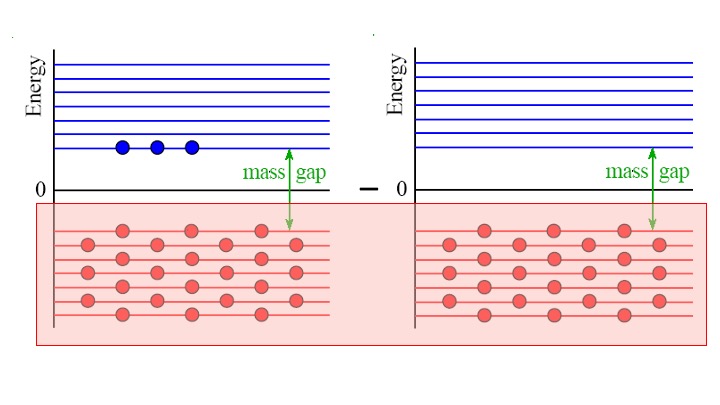}}\vspace{-2mm}
\caption{Schematic illustration of the calculation of the soliton mass, which is the sum
over the energies of the valence quarks, and the properly regularized sum over the sea quarks with vacuum
contribution subtracted, (\ref{eq:Msol1}). The upper panel corresponds to the configuration at the minimum.
In the limit of $r_0 \rightarrow 0$ (zero soliton size), shown in the lower panel, valence quarks go back
to the first positive energy level over the mass gap, and the sea is not polarized. Therefore, the sea contribution is canceled by 
the vacuum part (this cancellation is emphasized by the pink box).}%
\label{fig:solitonlevels}\vspace{-1mm}
\end{figure}

In the $\chi$QSM, the soliton mass is given as a sum over the  energies of the valence quarks and the sea quarks
 computed with respect to the vacuum and appropriately regularized (see \eg\ \cite{Goeke:2005fs})
\begin{equation}
M_{\rm sol}=N_c\left[ E_{\rm val}+\sum_{E_n<0} \left(E_n-E_n^{(0)}\right)\right]\,.
\label{eq:Msol1}
\end{equation}
This is schematically illustrated in the upper panel of Fig.~\ref{fig:solitonlevels}.

\begin{figure}[htb]
\centerline{%
\includegraphics[width=6.5cm]{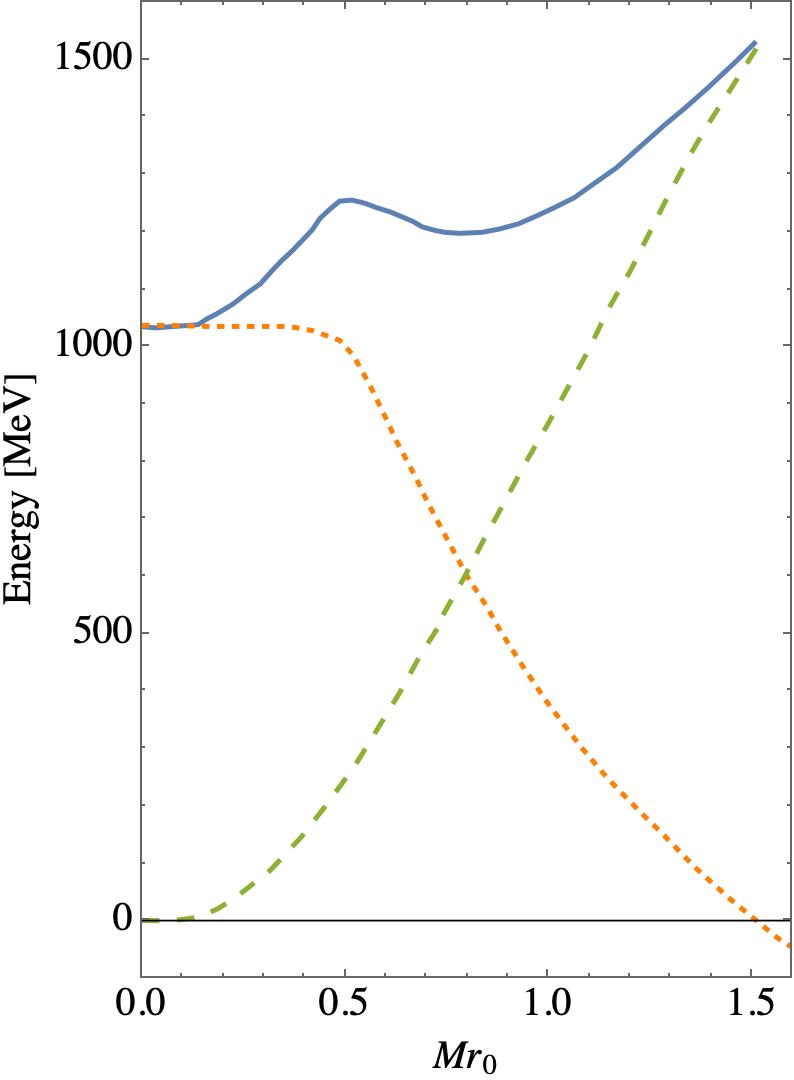}}
\caption{Soliton energy (mass) in MeV for $M=345$~MeV as a function of a dimensionless variational parameter $Mr_0$: solid (blue) --- total mass,
short-dashed (orange) --- energy of valence quarks, long-dashed (green) --- sea contribution. Minimum of $\sim1200$~MeV corresponds ro $r_0 \simeq 0.5$~fm.
Figure from Ref.~\cite{Diakonov:1988mg}.}%
\label{fig:solitonenergy}%
\end{figure}

Figure \ref{fig:solitonenergy} illustrates how the energy of the soliton changes with increasing $r_0$~\cite{Diakonov:1988mg}. We see that
at some small $r_0>0$, the valence energy levels fall into the mass gap and the energy of the Dirac sea is increasing,
however, the total energy is decreasing.
A stable configuration is reached for some $r_0^{\rm min}$. 
For $r_0> r_0^{\rm min}$, the sea energy starts wining, and the total energy increases.

As can be seen from Fig.~\ref{fig:solitonenergy}, in the limit $r_0 \rightarrow 0$, the sea energy goes to
zero, and the total energy is given by the valence levels only, see the lower panel of Fig.~\ref{fig:solitonlevels}. 
This is referred to as the  Non-Relativistic Quark Model
(NRQM) limit. In this limit, many quantities can be computed analytically. 
For $r_0 \rightarrow \infty$, the valence
level sinks into the Dirac sea, and the total soliton energy is given by the sum over the sea levels only. This
is called the Skyrme model limit. We can see from Fig.~\ref{fig:solitonenergy} that the true minimum is halfway between these two limits.

The advantage of the arctan Ansatz (\ref{Pr}) is best seen in the case of the Skyrme model where all integrals can be performed analytically.
Introducing the new dimensionless variable $x_0=Fe\,r_0$, the soliton mass takes the following form~\cite{Praszalowicz:1991aj}:
\begin{eqnarray}
M_{\rm sol} & = & \frac{F_{\pi}}{e} {\pi}^{2} \frac{3 \sqrt{2}}{16}
\left(4 x_{0} +\frac{15}{x_{0}}\right) \,. \label{eq:massMx} 
\end{eqnarray}
The term linear in $x_0$ comes from the Weinberg Lagrangian, whereas term $\sim 1/x_0$ from the Skyrme term. The minimum
is reached for $x_0=\sqrt{15/4}$.

Unfortunately, the soliton minimum energy $M_{\rm sol}\simeq1200 \div 1350$~MeV is much higher than the nucleon mass\footnote{In the SU(3)
case, this value should be compared with the mean octet mass $M_8=1154$~MeV.}. This is a common feature of  chiral
models including the Skyrme model~\cite{Adkins:1983ya,Praszalowicz:1985bt} and the chiral quark model as
well~\cite{Diakonov:1988mg}. The soliton mass scales like $M \sim N_c$, it is, however, a subject
to a major ${\cal O}(N_c^0)$ correction, which originates from
the quantum fluctuations of the meson field around
the classical soliton configuration. The corresponding
mass shift is called a Casimir energy
\cite{Moussallam:1991rj,Holzwarth:1992md} and is negative. Casimir corrections are typically 
ignored in phenomenological applications. As we will see, the mass splittings inside and between different baryon multiplets are
much better reproduced than absolute masses.

\subsection{SU(3) soliton and the collective quantization}

In the SU(3) case, the hedgehog Ansatz (\ref{eq:hedg}) is embedded in the  ``isospin corner''
(although other embeddings are also possible~\cite{Balachandran:1983dj,Balachandran:1985fb})
\begin{equation}
U_{0}=\left[
\begin{array}{ccc}
\multicolumn{2}{c}{\multirow{2}{*}{$u_0$}}    & 0 \\
\multicolumn{2}{c}{}  &0   \\
 0   & 0    & 1
\end{array} 
\right]\,.
\label{eq:Usu3}
\end{equation}
For the isospin embedding (\ref{eq:Usu3}), the static solution does not change. In order to provide the ``classical'' baryon with specific
quantum numbers, one has to consider an SU(3)-rotated pseudoscalar
field
\begin{equation}
U(t,\boldsymbol{r})=A(t)U_0(\boldsymbol{r})A^{\dagger}(t)
\label{eq:rotU}
\end{equation}
and derive the pertinent Lagrangian expressed in terms of the collective velocities $\mathrm{d}a_{\alpha}(t)/\mathrm{d}t$ defined as follows:
\begin{equation}
A^{\dagger}(t) \frac{\mathrm{d}A(t)}{\mathrm{d}t} = \frac{i}{2}  \sum_{\alpha=1}^{8}
\lambda_{\alpha}\frac{\mathrm{d}a_{\alpha}(t)}{\mathrm{d}t} \, .
\label{eq:adots}
\end{equation}
At this point, it is important to note that
$A \in\,$SU(3)/U(1)  rather than full SU(3), since for the hedgehog Ansatz~(\ref{eq:Usu3}), $ [\lambda_{8},U_{0}]\! =\! 0 $.
Therefore, matrix~$ A $ is defined up to a {\em local} U(1) factor
$ h= \exp(i \lambda_{8} \phi) $, \ie\ $ A $ and $Ah$ are equivalent. For this reason, the eighth coordinate $a_{8}(t)$
is not dynamical and does not appear in the kinetic energy of the rotating Skyrmion. Indeed, the collective
Lagrangian for the rotating hedgehog reads (for a review, see Ref.~\cite{Christov:1995vm})

\vspace{-5mm}
\begin{equation}
{\cal L}_{\rm coll}=-M_{\rm sol}+\frac{I_{1}}{2} \sum_{i=1}^{3} {\frac{\mathrm{d}a_{i}}{\mathrm{d}t}}^{2}
  +  \frac{I_{2}}{2} \sum_{k=4}^{7}{\frac{\mathrm{d}a_{k}}{\mathrm{d}t}}^{2}
  + \frac{N_{c}}{2 \sqrt{3}} \frac{\mathrm{d}a_{8}}{\mathrm{d}t}
  + \Delta m\,. \label{eq:Sklagra}
\end{equation}
The linear term in $\mathrm{d}a_8/\mathrm{d}t$ results in the constraint on the allowed Hilbert space.

One can see that (\ref{eq:Sklagra}) resembles the well-known quantum mechanical Lagrangian of a symmetric top.
Here, $M_{\rm sol}$ is the soliton mass discussed in the previous section, $I_{1,2}$ are moments
of inertia, and $\Delta m \sim m_s$ is the SU(3) symmetry-breaking piece, which we will treat
as a perturbation.

In order to construct the collective Hamiltonian, we have to perform the Legendre transformation, and --- even
more importantly --- identify the symmetries of (\ref{eq:Sklagra}) in order to associate collective momenta with
the generators of these symmetries. There are two symmetry groups which leave $ {\cal L}_{\rm coll} $
invariant

\vspace{-5mm}
\begin{eqnarray}
\begin{array}{llll}
A(t) & \rightarrow & 
g_{\rm L}\, A(t)\,,\qquad  & g_{\rm L}\in {\rm SU(3)_{L}}\,,\\
A(t) & \rightarrow & A(t) \,  g^{\dagger}_{\rm R}\,, & g_{\rm R}\in {\rm SU(2)_{R}}
\times {\rm U(1)}\,.
\end{array} \label{eq:gLgR}
\end{eqnarray}

Since $ A $ belongs to the coset space SU(3)/U(1) rather than to SU(3),
the {\em right} symmetry splits into the product of 
${\rm SU(2)_{R}\;and\;U(1)}$.
{\em Left} SU(3) symmetry corresponds to flavor, {\em right}
SU(2)  to  spin,  and 
 {\em right} U(1) factor results in the constraint~\cite{Guadagnini:1983uv,Jain:1984gp,Mazur:1984yf,Chemtob:1985ar}
\begin{equation}
Y' = \frac{N_c}{3}\,, \label{eq:YR}
\end{equation}
where $Y' $  is a hypercharge corresponding to the {\em right} U(1).

Wave functions for a quantum mechanical symmetric top are given in terms of Wigner $D$-functions~\cite{Landau1977}
$D_{ab}^{({\cal R})}(A)$, where ${\cal R}=(p,q)$ labels the SU(3) representation (in the case of a quantum mechanical
top  ${\cal R}$ is simply the angular momentum) and indices $a,b$ run over all states in representation ${\cal R}$.
Here, however, due to the constraint (\ref{eq:YR}) one  index runs only over states that have hypercharge equal to $Y'$.
This means that only representations  ${\cal R}$ that have states of $Y=Y'$ are allowed. In the present case, for
$N_c=3$, we have $Y'=1$ and the allowed representations are 
\begin{equation}
{\cal R}= {\bf 8},\, {\bf 10}, \, \overline{ {\bf 10}}, \, {\bf 27}, \, {\bf 35}, \, \overline{ {\bf 35}}, \ldots
\label{eq:SU3reps}
\end{equation}
We see that in addition to the octet and decuplet of positive-parity baryons, well known from the quark model, exotic representations, 
like $\overline{ {\bf 10}}$,
emerge, all of positive parity.

Skipping technicalities~\cite{Christov:1995vm}, the baryon wave function takes the following form\footnote{One can find different representations of
this wave function in the literature that are equivalent to the one used here.}:
\begin{equation}
\psi_{(B,\,J,J_{3})}^{(\mathcal{R})}(A)  
 = (-)^{J_{3}-Y^{\prime}/2}  \sqrt{\text{dim}(\mathcal{R})}\,
D_{(Y,\,T,\,T_{3}%
)(Y^{\prime},\,J,\,-J_{3})}^{(\mathcal{R})\ast}(A) \, .
\label{eq:rotwf}%
\end{equation}
Here, $B=(Y,T,T_{3})$ stands for the SU(3) quantum numbers of a baryon in question,
and the second index of the $D$ function, $(Y^{\prime},J,-J_{3})$, corresponds
to the soliton spin.

The pertinent rotational collective Hamiltonian takes the following form:
\begin{equation}
{\cal{H}}_{\rm rot}= M_{\rm{sol}}
+ \frac{1}{2I_{1}}J(J+1) +\;\frac{1}{2I_{2}}\left[
C_{2}({\cal{R}})-J(J+1)-\frac{3}{4}Y^{\prime2}\right] \, ,
\label{rotmass}%
\end{equation}
where $C_{2}(\mathcal{R})$ stands for the SU(3) Casimir operator and possible $\mathcal{R}$s are given by (\ref{eq:SU3reps}).
Note that the last term proportional to $Y^{\prime 2}$ cancels out the last term in $C_{2}(\mathcal{R})$ and therefore,
as mentioned above, the rotational Hamiltonian does not depend explicitly on $Y^{\prime}$.

The collective Hamiltonian and constraint (\ref{eq:YR}) are exactly the same in the chiral quark model and in the Skyrme model.
The only obvious difference is that the soliton mass and the moments of inertia are in the Skyrme model expressed in terms of space
integrals over some functionals of the profile function $P(r)$, while in the case of the quark model, they are given as regularized sums
over the one-particle energy levels of the Dirac Hamiltonian corresponding to~(\ref{chiQM}).

\newpage

\subsection{Mass splittings and decay widths}
\label{ssec:mass_width}

In the Skyrme model the only term responsible for the nonzero meson ($=\mathrm{quark}$) masses is given by (\ref{eq:mlagr}).
For the rotating $U$ field (\ref{eq:rotU}), the first term proportional to $a$ gives merely a constant, whereas the second term is
proportional to
\begin{equation}
{\rm Tr} \left(  \left(U_0+U_0^{\dagger}\right) A^{\dagger}\lambda_8 A   \right) \, .
\end{equation}
Using the identity
\begin{equation}
 A^{\dagger}\lambda_a A =D^{(\boldsymbol{8})}_{ab}(A)\lambda_b 
\end{equation}
and the properties of the hedgehog Ansatz, we get that the symmetry breaking Hamiltonian is given by
\begin{equation}
{\cal H}_{\rm br}=\alpha D^{(\boldsymbol{8})}_{88}(A) \, .
\label{eq:HbrSkyrme}
\end{equation}
Here, $\alpha$ is a known functional of the profile function $P$ proportional to $N_c$. Index 8 of the Wigner $D$ function 
corresponds to $8\!=\!(Y\!=\!0,T\!=\!0,T_3\!=\!0)$.
For massless light quarks $u$ and $d$, the coefficient $\alpha$ is proportional to $m_s\sim m_K^2$.

In the chiral quark model (\ref{chiQM}), the systematic expansion in rotational velocities yields new terms not present in the Skyrme model~\cite{Christov:1995vm}
\begin{equation}
 m_s D^{(\boldsymbol{8})}_{8a}(A)\; K_{ab}\frac{\mathrm{d}a_b(t)}{\mathrm{d}t}
\label{eq:Kab}
\end{equation}
where $K_{ab}$ denotes the tensor of the anomalous moments of inertia, which originate from the anomalous (imaginary) part of
the Euclidean quark model action. While $K_{ab}\sim N_c$, the rotational velocities are ${\mathrm{d}a_b(t)}/{\mathrm{d}t}\sim 1/N_c$ (that is because
velocities are proportional to collective momenta divided by moments of inertia, which are ${\cal O}(N_c)$). One would, therefore,
naively expect that corrections (\ref{eq:Kab}) are of the order ${\cal O}(m_s N_c^0)$, while the leading term (\ref{eq:HbrSkyrme}) is of the
order ${\cal O}(m_s N_c)$.

With these new terms, the full symmetry-breaking Hamiltonian is of the form~\cite{Diakonov:1997mm}
\begin{equation}
{\cal H}_{\mathrm{{br}}}=\alpha\,D_{88}^{(\boldsymbol{8})}(A)+\beta\,\hat{Y}+\frac{\gamma}{\sqrt{3}%
}\sum_{i=1}^{3}D_{8a}^{(\boldsymbol{8})}(A)\,\hat{J}_{a}\,, \label{eq:Hsb}%
\end{equation}
where $\alpha$, $\beta$, and $\gamma$ are proportional to the strange
quark mass. Furthermore,
$\alpha$  scales as $N_c$, and $\beta$ and $\gamma$
scale as $N_c^0$. $\hat{Y}$ and $\hat{J}_{a}$ are hypercharge and spin operators, respectively.

In the large-$N_c$ limit, baryons consist of $N_c$ quarks and, therefore, the hypercharge eigenvalue of the physical states
is also $Y\sim N_c$. This means that the second term in (\ref{eq:Hsb}), including $\hat{Y}$, is of the order ${\cal O}(m_s N_c)$ 
like~(\ref{eq:HbrSkyrme})\footnote{Matrix elements of $D_{88}^{(\boldsymbol{8})}$ are ${\cal O}(N_c^0)$.}.
It was Gudagnini~\cite{Guadagnini:1983uv} who argued that $\beta\,\hat{Y}$ should be added to (\ref{eq:HbrSkyrme}) in the Skyrme model. 
In the chiral quark model, it arises naturally
from the gradient expansion.

Since we have identified the symmetries of the soliton, it is straightforward to compute the pertinent currents, in particular, the axial
current~\cite{Blotz:1994wi}. 
The axial current is of interest here, since via the Goldberger--Treiman relation it can be related to strong baryon decays\footnote{This
approach to the width calculations has been criticized  in the literature, see \eg\ Ref.~\cite{Walliser:2005pi}.}.
In the nonrelativistic limit for the initial and final baryons, $B_1$ and $B_2$ respectively,
the baryon--baryon--meson coupling can be written in the following form:
\begin{equation}
\mathcal{O}_{\varphi}=3 \sum_i \left[  G_{0}D_{\varphi\,i}^{(\boldsymbol{8})}-G_{1}\,d_{ibc}%
D_{\varphi\,b}^{(\boldsymbol{8})}\hat{S}_{c}-G_{2}\frac{1}{\sqrt{3}}D_{\varphi\,8}%
^{(\boldsymbol{8})}\hat{S}_{i}\right]  \,\frac{p_{i}}{M_{1}+M_{2}}\,,  \label{eq:dec-op}%
\end{equation}
where $M_{1,2}$ denote masses of the
initial and final baryons and $p_{i}$ is the
c.m. momentum of the outgoing meson, denoted as $\varphi$, of mass 
$m$
\begin{equation}
\left\vert \boldsymbol{p}\, \right\vert =p=\frac{1}{2M_{1}}\sqrt{\left(M_{1}^{2}%
-(M_{2}+m)^{2}\right)\left(M_{1}^{2}-(M_{2}-m)^{2}\right)} \, .
\end{equation}
The factor of 3 in Eq.~(\ref{eq:dec-op}) is a matter of convenience because it cancels in the averaged square
of ${\cal O}_{\phi}$, and the factor of $M_1+M_2$ is a matter of choice (see below).

The decay width is related to the matrix element of
$\mathcal{O}_{\varphi }$ squared, summed over the final, and averaged
over the initial spin and isospin  denoted as $\overline{\left[ 
\ldots\right]  ^{2}}$,  see Appendix of
Ref.~\cite{Diakonov:1997mm} for details of the corresponding
calculations
\begin{equation}
{\mit\Gamma}_{B_{1}\rightarrow B_{2}+\varphi}=\frac{1}{2\pi}\overline{\left\langle
B_{2}\left\vert \mathcal{O}_{\varphi}\right\vert B_{1}\right\rangle ^{2}%
}\,\frac{M_{2}}{M_{1}}p\,.
\label{eq:Gamma_phi}
\end{equation}
Factor $M_{2}/M_{1}$, used already in Ref.~\cite{Kim:2017khv}, is the same as in heavy baryon chiral perturbation theory (HBChPT); see \eg\ Refs.~\cite{Cheng:2006dk,Cheng:2015dk1}. 

Here, some remarks are in order. While the mass spectra are given as systematic
expansions  both in $N_c$ and  $m_s$,  the decay widths cannot be organized in a similar way. 
They depend on modeling and `educated' guesses, and hence are subject to additional
uncertainties~\cite{Ellis:2004uz}. The most important uncertainty comes from the fact that the baryon masses $M_1$ and $M_2$
are formally infinite series in $N_c$ and $m_s$. 
The same holds for the momentum of the outgoing meson. It is a common practice
to treat the phase factor exactly rather than expand it up to a given order in $N_c$
and $m_s$, despite the fact that in ${\cal O}_{\varphi}$, only a few first terms in $1/N_c$
and $m_s$ are included.  Here, we have adopted
a convention with $M_1+M_2$ in (\ref{eq:dec-op}) and $M_2/M_1$ in (\ref{eq:Gamma_phi}), for other choices, see \eg\ \cite{Ellis:2004uz}.
Formally, in the large-$N_c$ limit and small-$m_s$ limit, $M_1=M_2$ and both conventions are identical. Nevertheless, if we 
use physical masses for $M_{1,2}$, different conventions will result in different numerical results.

The leading term proportional to $G_0\sim N_c$ was introduced already in the Skyrme model in Ref.~\cite{Adkins:1983ya},
whereas the subleading terms $G_{1,2}\sim N_c^0$ were
derived in the chiral quark model~\cite{Diakonov:1997mm,Blotz:1994wi}.

Since we know the collective wave functions (\ref{eq:rotwf}), it is relatively straightforward to compute the matrix elements
for the mass splittings and decay widths. They are simply given in terms of the SU(3) Clebsch--Gordan coefficients~\cite{deSwart:1963pdg}.

\subsection{Heavy baryons}
\label{ssec:HBs}

In the
quark model, a heavy baryon consists
of a heavy quark and two light quarks. When the mass of the heavy quark
 $m_Q\rightarrow \infty$,  the spin of the heavy quark $\boldsymbol{
 S}_Q$ is conserved, which indicates that the spin of the light-quark
degrees of freedom is also conserved: $\boldsymbol{S}_{\mathrm{L}} \equiv
\boldsymbol{S}-\boldsymbol{S}_Q$~\cite{Isgur:1989vq, Isgur:1991wq,
Georgi:1990um}. Due to this heavy-quark spin symmetry, the 
total spin of the light quarks can be considered as a good
quantum number. This suggests that in the first
approximation, a heavy baryon can be viewed as the bound state of a heavy
quark and a diquark.

In the large-$N_c$ limit, heavy baryons consist of a heavy quark
and $N_c-1$ light quarks rather than a diquark. In this limit,
the $N_c-1$ valence quarks produce the mean field which hardly
differs from the one produced by $N_c$ quarks.
Therefore, the ``diquark'' system can be described as a quark-soliton
in close analogy to the light baryons~\cite{Diakonov:2010tf}.
Indeed, when the mass of one quark is included\footnote{Note that the soliton is formally calculated in
the chiral limit, where the current quark masses are equal to zero.} 
and the limit 
$m_Q \rightarrow \infty$ is formally performed, then the soliton energy (\ref{eq:Msol1}) reads as follows:
\begin{eqnarray}
M_{\rm sol}&=&(N_c-1)\left[ E_{\rm val}+\sum_{E_n<0} \left(E_n-E_n^{(0)}\right)\right]  \notag \\
&&+\left[ E_{\rm val}(m_Q)+\sum_{E_n<0} \left(E_n(m_Q)-E_n^{(0)}(m_Q)\right)\right]\,.
 \label{eq:Msol2}                  
\end{eqnarray}
It was argued in Ref.~\cite{Goeke:2005fs} that for large $m_Q$,  the sum
over the sea quarks in the second
line of Eq.~(\ref{eq:Msol2}) vanishes, and $E_{\rm val}(m_Q)\approx m_Q$. One copy of the
soliton ceases to exist; however, the remaining $N_c-1$ quarks still form a stable soliton.

In the ``diquark'' case,
the constraint (\ref{eq:YR}) is modified $Y'=(N_c-1)/3$, and the lowest allowed
representations are $\overline{\bf{3}}$ and $\bf{6}$ of spin 0 and spin 1, respectively, exactly as in the
quark model. Adding a heavy quark, one gets one antitriplet and two sextets of the total spin 1/2 and 3/2.
Therefore, one has
to introduce a spin--spin interaction~\cite{Zeldovich} to remove
spin $1/2$ and $3/2$ degeneracy of the sextet states. The hyperfine
coupling --- the only parameter undetermined from the light sector ---
has to be fixed from the experimental data.

This program was successfully carried over in Refs.~\cite{Kim:2017khv,Yang:2016qdz,Polyakov:2022eub}.
Apart from regular baryons, the model predicts exotic heavy baryons belonging to $\overline{\boldsymbol{15}}$ representaion of SU(3) flavor of spin 1/2 and 3/2~\cite{Kim:2017khv,Kim:2017jpx,Praszalowicz:2022hcp}. Possible candidates for exotic charm baryons
are two (out of five) recently discovered by the LHCb~\cite{Aaij:2017nav,LHCb:2021ptx} and confirmed by  Belle~\cite{Belle:2017ext}
 ${\mit\Omega}_c^{(0)}$ states, namely ${\mit\Omega}_c^{(0)}(3050)$ and ${\mit\Omega}_c^{(0)}(3119)$.

\section{Exotic phenomenology}
\label{sec:pheno}

\subsection{${\Theta^+}$ mass}
\label{ssec:Thetamass}

Let us summarize the results of Section~\ref{sec:effQCD}. In the chiral limit, the mass formula for baryons resembles the one of
a quantum mechanical symmetric top with a constraint that selects allowed SU(3) representations (\ref{eq:SU3reps}). We will
be mostly interested in the exotic antidecuplet. Mass splittings between different multiplets are related to the moments of inertia
$I_{1,2}$ of the rotating soliton
\begin{equation}
{\mit\Delta}_{{\bf 10}-{\bf 8}}=\frac{3}{2} \frac{1}{I_1} \, ,\qquad {\mit\Delta}_{\overline{\bf 10}-{\bf 8}}=\frac{3}{2} \frac{1}{I_2}\, .
\label{eq:10-8split}
\end{equation}
We see that $I_2$, which is absolutely necessary for predicting the masses of exotic baryons, cannot be 
constrained by experimental data on ordinary baryons. The same is true for the symmetry-breaking terms 
and the decay operator\footnote{This is true in the first order of the perturbation theory.}.

Chiral symmetry-breaking terms following from the fact that $m_s>m_{u,d}$ generate mass splittings within the SU(3)$_{\rm flavor}$ multiplets~\cite{Diakonov:1997mm}
\begin{eqnarray}
\Delta M_{\bf 8}&=&\frac{1}{20}\left(2\alpha+3\gamma\right) + \frac{1}{8}\,\left[\left(2\alpha+3\gamma\right)+4\left(2\beta-\gamma\right) \right] \, Y  \notag \\
 &&-\frac{1}{20}\left(2\alpha+3\gamma\right) \left[ T(T+1)-\frac{1}{4}Y^2 \right] \, , \notag \\
\Delta M_{\bf 10}&=&\frac{1}{16}\,\left[\left(2\alpha+3\gamma\right)+8\left(2\beta-\gamma\right)\right]\, Y\, ,\notag \\
\Delta M_{\overline{\bf 10}}&=&\frac{1}{16}\,\left[\left(2\alpha+3\gamma\right)+8\left(2\beta-\gamma\right)+4\gamma \right]\,Y \, ,
\label{eq:massabg}
\end{eqnarray}
where parameters $\alpha,\, \beta$, and $\gamma$ are proportional to $m_s-m_{u,d}$. Note that in the Skyrme model, $\gamma=0$ and $\beta=0$ if
we do not take into account the Guadagnini term~\cite{Guadagnini:1983uv}. Equations (\ref{eq:massabg}) are written in a form, 
from which one can immediately see that the
mass splittings of nonexotic baryons depend in fact only on two combinations of parameters $\alpha,\, \beta$, and $\gamma$, namely
on $2\alpha+3\gamma$ and $2\beta-\gamma$, whereas the
mass splittings in exotic $\overline{\bf 10}$ depend additionally on~$\gamma$. 
This means that we cannot predict mass splittings within $\overline{\bf 10}$ 
from the spectrum of nonexotic baryons.

We can, of course, compute multiplet splittings (\ref{eq:10-8split}) and mass splittings (\ref{eq:massabg}) in some specific model. A good example is
decuplet --- octet splitting ${\mit\Delta}_{{\bf 10}-{\bf 8}}\simeq 230~{\rm MeV}$.
In the Skyrme model with the arctan Ansatz (\ref{Pr}),  moments of inertia take a very simple form~\cite{Praszalowicz:1991aj}
\begin{eqnarray}
I_{1} &=& \frac{1}{e^{3}F_{\pi}}\pi^{2}\frac{\sqrt{2}}{12}\,\left(6x_{0}^3+25x_{0}\right)\,,
\notag \\
I_{2} &=& \frac{1}{e^{3}F_{\pi}}\pi^{2}\frac{\sqrt{2}}{16}\,\left(4x_{0}^3+9x_{0}\right)\,,
\end{eqnarray} 
where $x_0=\sqrt{15/4}$ is obtained by minimizing the soliton mass (\ref{eq:massMx}). One finds that for $F=186$~MeV, decuplet--octet splitting
${\mit\Delta}_{{\bf 10}-{\bf 8}}$ of Eq.~(\ref{eq:10-8split}) is reproduced for $e\simeq 4.45$, well within the expected
range. 

Here, we are interested in the first exotic representation, namely $\overline{\bf 10}$ depicted in Fig.~\ref{fig:mixing}. 
Once $e$ is fixed, we can compute antidecuplet--octet splitting
\begin{equation}
{\mit\Delta}_{\overline{\bf 10}-{\bf 8}}=\frac{3}{2} \frac{1}{I_2} \simeq 600~{\rm MeV} \, .
\label{eq:10bar-8split}
\end{equation}
This is much less than the naive quark model expectations~\cite{Biedenharn:1984qg} and agrees with an old estimate of 
Ref.~\cite{Biedenharn:1984su} that led its authors 
to conclude: \textit{Since the theory is a low energy effective theory, we believe that this
gives an aposteriori excitation energy limit on the validity}. Indeed, rigidly rotating soliton predicts an infinite tower of exotic
representations (\ref{eq:SU3reps}), and it is clear that this picture has to break down at some point. The question is: does it break already
for $\overline{\bf 10}$\,?

Although the formulas for the mass splittings and decay couplings have been derived in some specific models, their general form
is to a large extent model-independent, as it follows from the hedgehog symmetry. This observation led Adkins and Nappi~\cite{Adkins:1984cf}
to extract moments of inertia and other quantities directly from the data rather than computing them in some model. Here,
as mentioned above, we immediately
encounter a problem, since we have no handle on the $I_2$ moment of inertia, as it does not enter into any formula for nonexotic
baryons. Similarly, parameter $\gamma$ cannot be constrained from the nonexotic baryons alone (\ref{eq:massabg}).

One can, however, make a rough estimate of the ${\mit\Theta}^+$ mass assuming Skyrme model ${\mit\Delta}_{\overline{\bf 10}-{\bf 8}}$
value (\ref{eq:10bar-8split}) and observing that the mass splittings in $\overline{\bf 10}$ are approximately equal to the ones in
the decuplet, $140 \div 150$~MeV. One then obtains that ${\mit\Theta}^+$ mass is as low as $\sim1460$~MeV~\cite{Praszalowicz:2003ik}. 
More detailed analyses
in the Skyrme model~\cite{Praszalowicz:2003ik,Praszalowicz:1987em} and in the quark-soliton model~\cite{Diakonov:1997mm} 
led to the mass $1530 \div 1540$~MeV, which was reinforced by the experimental
results of LEPS~\cite{LEPS:2003wug} and DIANA~\cite{DIANA:2003uet}.

Interestingly, the mass of another truly exotic pentaquark state, namely ${\mit\Xi}_{3/2}$ (see Fig.~\ref{fig:mixing}), was estimated
$M_{{\mit\Xi}_{3/2}}\simeq 1785$~MeV in the Skyrme model\break \cite{Praszalowicz:2003ik,Praszalowicz:1987em}, and
$M_{{\mit\Xi}_{3/2}}\simeq 2070$~MeV in the quark-soliton model~\cite{Diakonov:1997mm}. In 2004, the NA49 Collaboration
at CERN reported an observation of $S=-2$ and $Q=-2$ exotic baryon at 1862~MeV~\cite{NA49:2003fxh}. However,
a more recent analysis of NA61/SHINE~\cite{NA61SHINE:2020mti} did not confirm the NA49 result and no peak
corresponding to ${\mit\Xi}_{3/2}$ in ${\mit\Xi}+\pi$ spectra in the mass range $1700 \div 2400$~MeV was found.
We will further discuss this in Section~\ref{ssec:a10}.

\subsection{$\Theta^+$ decay width}
\label{ssec:Thetawidth}

Calculations of the pentaquark decay widths retaining only the first leading term in (\ref{eq:dec-op}) 
yield results that are of the same order 
as the width of ${\mit\Delta}$, namely $\sim 100$~MeV~\cite{Diakonov:1997mm,Weigel:1998vt}. 
It is, therefore, essential to include
the subleading terms $G_1$ and $G_2$ from Eq.~(\ref{eq:dec-op}) in order to account for the small width of ${\mit\Theta}^+$.

There are two possible strategies to constrain the decay parameters $G_{0,1,2}$: 
one can either try to employ directly data on strong decays,
or use the Goldberger--Treiman relation
\begin{equation}
\left\{  G_{0},G_{1},G_{2}\right\}  =\frac{M_{1}+M_{2}}{2F_{\varphi}}\frac
{1}{3}\left\{  a_{0},-a_{1},-a_{2}\right\}\,,
\label{eq:GTrelation}
\end{equation}
where constants $a_{0,1,2}$ enter the definition of the axial-vector
current~\cite{Blotz:1994wi,Praszalowicz:1998jm,paradox} and can be extracted from the
semileptonic decays of  the
  baryon octet~\cite{Yang:2015era}. The relations of the constants
$a_{0,1,2}$ to the nucleon axial charges in the chiral limit read as follows:  
\begin{align}
g_A^{(0)}&=\frac12 a_2 \, , \notag \\
g_{A}^{(3)}&=\frac{7}{30}\left( -a_{0}+\frac{1}{2}a_{1}+\frac{1}{14}a_{2}\right) \, , \notag \\
g_{A}^{(8)}&=\frac{1}{10 \sqrt 3}\left(
-a_{0}+\frac{1}{2}a_{1}+\frac{3}{2}a_{2}\right) \, .
\label{eq:axialc}
\end{align}
The final formula for the decay width in terms of the axial-vector
constants $a_{0,1,2}$ takes the following form~\cite{Diakonov:1997mm}:%
\begin{align}
\label{eq:Gammageneral}
{\mit\Gamma}_{B_{1}\rightarrow B_{2}+\varphi}  &  \sim \frac{p^{3}%
}{F_{\varphi}^{2}}\frac{M_{2}}{M_{1}}G_{\mathcal{R}_{1}\rightarrow
\mathcal{R}_{2}}^{2} \, .
\end{align}
Here, $\mathcal{R}_{1,2}$ are the SU(3) representations of the initial and final baryons and 
the omitted proportionality factor contains the SU(3) isoscalar factors and the ratio of dimensions of representations 
$\mathcal{R}_{1,2}$  (see \eg\ Eq.~(8) in Ref.~\cite{Kim:2017khv}).
The decay constants $G_{\mathcal{R}_{1}\rightarrow
\mathcal{R}_{2}}$ are calculated from the matrix elements of
(\ref{eq:dec-op})  and read as follows: 
\begin{equation}
G_{10\rightarrow8}=-a_0+\frac{1}{2}a_1\,,\qquad G_{\overline{10}\rightarrow 8}=-a_0-\frac{N_c+1}{4}a_1-\frac{1}{2}a_2 \, ,
\label{eq:G_vs_a}
\end{equation}
where we have explicitly displayed the $N_c$ dependence following from the pertinent $N_c$ dependence of the flavor SU(3)
Clebsch--Gordan coefficients~\cite{Praszalowicz:2003tc}.

As we discussed in Section~\ref{ssec:hedgehog}, for small soliton size (the so-called Non-Relativisic Quark Model limit, NRQM), 
one can compute constants $a_{0,1,2}$
analytically~\cite{Praszalowicz:1998jm}
\begin{equation}
a_{0}\rightarrow-(N_{c}+2)\,,\qquad a_{1}\rightarrow4,\;a_{2}\rightarrow2 \, .
\label{eq:NRa123}%
\end{equation}
The reader may convince herself/himself that in this limit (for $N_c=3$), 
\begin{equation}
g_{A}
\rightarrow\frac{5}{3}\,,
\end{equation}
which is equal to the naive quark model result for $g_{A}$. In this limit,
\begin{equation}
G_{10\rightarrow8}=N_c+4\,,\qquad G_{\overline{10}\rightarrow 8}=0 \, .
\label{eq:QMlimit4G}
\end{equation}
We see that the decay constant of antidecuplet is zero! The cancellation takes place for any $N_c$~\cite{Praszalowicz:2003tc}.
 This explains the smallness of ${\mit\Theta}^+$ width, which
for the realistic soliton size, is not equal to zero, but still very small (see below). In contrast, the decuplet 
decay constant is large explaining the large width of ${\mit\Delta}$ resonance. For the $N_c$ dependence 
of the decay widths including the phase-space factor $p^3$, see Ref.~\cite{Praszalowicz:2003tc}.

\begin{figure}[htb]
\centerline{%
\includegraphics[width=7.8cm]{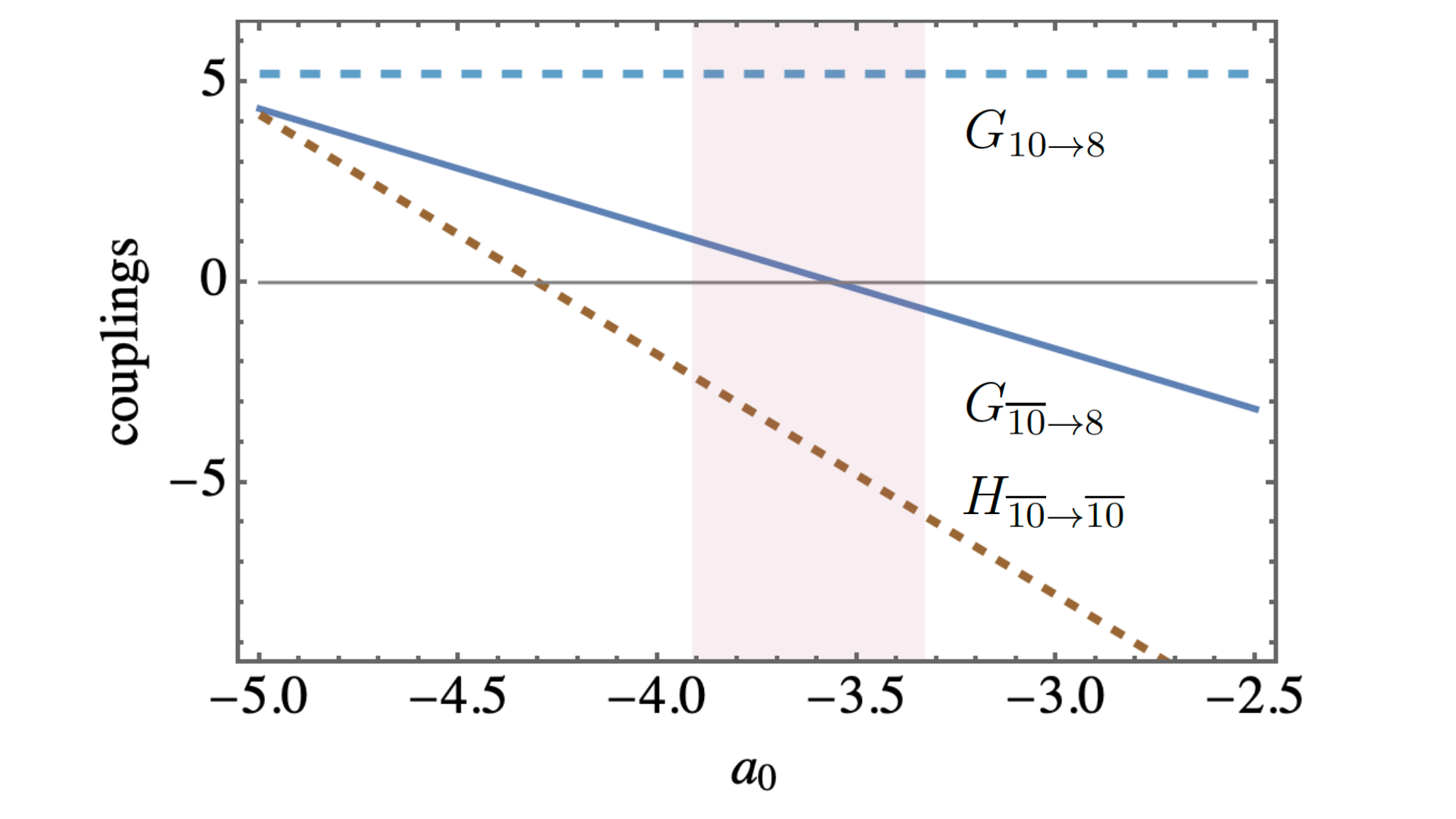}}\vspace{-2mm}
\caption{Couplings $G_{10\rightarrow 8}$ (upper long-dashed line), $G_{\overline{10}\rightarrow 8}$ (middle solid line),
and $H_{\overline{10}\rightarrow \overline{10}}$ (lower short-dashrd line)
as functions of $a_0$. The shaded area corresponds to the NJL model range~\cite{Blotz:1994wi}.}%
\label{fig:couplings}
\end{figure}

Unfortunately, it is not possible to extract all three couplings $a_{0,1,2}$ from the axial decays of hyperons, since they depend
only on the linear combination $-a_0+a_1/2$ and $a_2$ (\ref{eq:axialc}), while for $G_{\overline{10}\rightarrow 8}$,
we need all three of them separately. To get some insight into the numerical value of
$G_{\overline{10}\rightarrow 8}$, we can use experimental data $g_A^{(3)}=1.25$ and $g_A^{(0)}=0.24$ (\ref{eq:axialc}) yielding
\begin{equation}
-a_0+\frac{1}{2} a_1=5.21\,,\qquad a_2=0.48\,.
\label{eq:fromfit}
\end{equation}
Note that the first entry in Eq.~(\ref{eq:fromfit}) is equal to $G_{10\rightarrow8}$ (\ref{eq:G_vs_a}).
From Eq.~(\ref{eq:fromfit}), one can predict  $g_A^{(8)}=0.34$ in good agreement with the experimental value of 0.31. Now,
we can compute $a_1$ as a function of $a_0$ (which is negative) and plot   $G_{10\rightarrow 8}$ and 
$G_{\overline{10}\rightarrow 8}$. This is shown 
in Fig.~\ref{fig:couplings} where we also display the shaded area corresponding to the NJL 
model calculations~\cite{Blotz:1994wi}
for different constituent  quark masses $M$. 
We see that $G_{\overline{10}\rightarrow 8}$ is for a wide range of $a_0$ much smaller than $G_{10\rightarrow8}$
(including zero for $a_0=-3.55$). A rather involved fit to the hyperon decays with $m_s$ corrections included\footnote{Including $m_s$ corrections allows to
disentangle all three couplings $a_{0,1,2}$.} of Ref.~\cite{Yang:2015era} gives $a_0=-3.51$ corresponding to $G_{\overline{10}\rightarrow 8}=-0.23$.
In other words, $(G_{10\rightarrow8}/G_{\overline{10}\rightarrow 8})^2\sim 500$.
However, this result is strongly model-dependent and subject to unknown systematic uncertainty. 

In any case, the message from this consideration is clear: the ${\mit\Theta}^+$ decay width  is small irrespectively of the
prefactors entering Eq.~(\ref{eq:Gammageneral}). Let us remind that a misprint in a prefactor for the ${\mit\Delta}$ decay in Eq.~(42) 
of Ref.~\cite{Diakonov:1997mm} from 1997, triggered in 2004 a discussion~\cite{Jaffe:2004qj,Diakonov:2004ai,Jaffe:2004dc}
about the width of ${\mit\Theta}^+$, which was originally estimated to be $15$~MeV. This anyway a relatively large width followed from a rather
conservative estimate of $a_0$, which --- as can be seen from Fig.~\ref{fig:couplings} --- has no impact on the ${\mit\Delta}$ decay width
(\ie\  on $G_{10\rightarrow 8}$).
Today, it is clear that the ${\mit\Theta}^+$ width must be much
smaller, presumably below $0.5$~MeV.
 
Given the fact that the chiral limit $g_{{\mit\Theta} NK}=G_{\overline{10}\rightarrow 8}$ is very small, chiral symmetry-breaking 
effects are of importance. There are two kinds of $m_s$ corrections: corrections to the decay operator ${\cal O}_{\varphi}$
(\ref{eq:dec-op}) and the wave function mixing. Corrections to ${\cal O}_{\varphi}$ are rather complicated introducing five 
new terms~\cite{Blotz:1994wi} and we will not discuss them here. On the contrary, the wave function mixing is relatively
easy to estimate~\cite{Praszalowicz:2004dn}. Indeed, cryptoexotic members of antidecuplet can mix with the ground-state octet, and the mixing
angle will be small due to the fact that Gell-Mann--Okubo mass formulas are very well satisfied, leaving little space
for mixing. Nevertheless, the symmetry-breaking Hamiltonian (\ref{eq:Hsb}) inevitably introduces the representation mixing, which
in the case of the nucleon, takes the following form:
\begin{equation}
\mid N^{\rm phys}\, \rangle =\cos\alpha \mid N_{\bf 8}\, \rangle+ \sin\alpha \mid N_{\overline{\bf 10}}\, \rangle \,,
\label{eq:Nmixing}
\end{equation}
where $\sin \alpha >0$ is small and therefore $\cos \alpha \simeq 1$. Note that ${\mit\Theta}^+$ does not mix, and 
$\mid {\mit\Theta}^{\rm phys}\, \rangle = \mid {\mit\Theta}_{\overline{\bf 10}}\, \rangle$. Therefore, the decay of ${\mit\Theta}^+$ to
$KN$ proceeds either directly to $\mid N_{\bf 8} \rangle$ or through mixing with $\mid N_{\overline{\bf 10}} \rangle$,
leading to a new decay constant $H_{\overline{10}\rightarrow \overline{10}}$
\begin{equation}
\label{eq56}
g_{{\mit\Theta} NK}\simeq G_{\overline{10}\rightarrow 8} +\sin \alpha \, H_{\overline{10}\rightarrow \overline{10}}\,,
\end{equation}
where~\cite{Praszalowicz:2004dn}

\vspace{-8mm}
\begin{equation}
H_{\overline{10}\rightarrow \overline{10}}=-a_0-\frac{5}{2} a_1 + \frac{1}{2} a_2 \, .
\end{equation}

We plot $H_{\overline{10}\rightarrow \overline{10}}$ in Fig.~\ref{fig:couplings}.~We see that 
in absolute value,
it is larger than 
$G_{\overline{10}\rightarrow 8} $, and in a wide range of $a_0$, has the opposite sign, leading to a further
suppression of $g_{{\mit\Theta} NK}$. When discussing decay widths of other members of antidecuplet, it is quite natural to include
other mixing patterns that go beyond the present model. We discuss one such possibility in the next section. 

\vspace{-3mm}
\subsection{Exotic antidecuplet}
\label{ssec:a10}

The existence of ${\mit\Theta}^+$ implies the existence of all members of antidecuplet, see Fig.~\ref{fig:mixing}. As explained in
Section~\ref{ssec:mass_width}, one cannot constrain the masses of the remaining members of $\overline{\boldsymbol{10}}$
using as input masses of nonexotic baryons, as we have no handle on the strange moment of inertia $I_2$ (\ref{rotmass}) 
and the splitting parameter  $\gamma$ (\ref{eq:Hsb}). Definitely, apart from the ${\mit\Theta}^+$ mass, one needs yet another input.
In the pioneering work~\cite{Diakonov:1997mm}, the situation was similar, although the goal was to predict 
the ${\mit\Theta}^+$ mass.

One possible input is the pion--nucleon ${\mit\Sigma}_{\pi N}$ term related to the combination of parameters 
$\alpha$ and $\beta$~\cite{Diakonov:1997mm}
\begin{equation}
{\mit\Sigma}_{\pi N}=-\frac{3}{2}\frac{m_u+m_d}{m_s}\left(\alpha+\beta\right) \, ,
\label{eq:Sigma_term}
\end{equation}
which is linearly-independent of the combinations entering the mass splittings  (\ref{eq:massabg}).
Unfortunately, the experimental value of the ${\mit\Sigma}_{\pi N}$ term varied over the years from $\sim 40$ to 
$\sim 80$~MeV~\cite{Pavan:2001wz,Alarcon:2021dlz}
being, therefore,
rather useless for the precise determination of the antidecuplet masses. Moreover, the ratio of the current quark masses in (\ref{eq:Sigma_term})
is subject to $\sim 25\%$ error~\cite{Ellis:2004uz}.

As already mentioned at the end of Section~\ref{ssec:Thetamass},
in 2003, the NA49 Collaboration at CERN announced the observation of an exotic ${\mit\Xi}^{--}$ pentaquark (lower left vertex in Fig.~\ref{fig:mixing}) at
1.862~GeV~\cite{NA49:2003fxh}. If confirmed, it would be the second input besides ${\mit\Theta}^+$ to anchor the exotic antidecuplet.
Unfortunately, 17 years later, the successor of NA49, the NA61/SHINE Collaboration, did not confirm the ${\mit\Xi}^{--}$ peak around
1.8~GeV with 10 times greater statistics~\cite{NA61SHINE:2020mti}. One possible reason for this nonobservation might be the  extremely
small width of ${\mit\Xi}^{--}$. Indeed, in Ref.~\cite{Ellis:2004uz}, it was argued that in the SU(3) symmetry limit (\ie\  without mixing effects),
this width is up to a factor of $\sim 2$ equal to the width of ${\mit\Theta}^+$, \ie\ of the order of 1~MeV. The original analysis of NA49~\cite{NA49:2003fxh} 
reported the width ${\mit\Xi}^{--}$ below detector resolution of 18~MeV, while NA61/SHINE~\cite{NA61SHINE:2020mti} does not discuss
their sensitivity to the width of~${\mit\Xi}^{--}$.

There exists, however, a potential candidate for a cryptoexotic pentaquark, namely the nucleon resonance $N(1685)$~\cite{Kuznetsov:2008ii}, which was
initially announced  by the GRAAL Collaboration at the NSTAR Conference  in 2004~\cite{GRAAL:2004ndn}. $N(1685)$ was observed 
in the quasi-free neutron cross-section and in the $\eta n$ invariant mass spectrum~\cite{GRAAL:2006gzl,Kuznetsov:2008hj}, 
and was later confirmed by other groups:
CBELSA/TAPS~\cite{CBELSA:2008epm} and LNS-Sendai~\cite{Shimizu:2008cwq}. 
We refer the Reader to the article by Strakovsky~\cite{Strakovsky:2024ppo} in this volume to learn
more about $N(1685)$.
The observed structure 
can be interpreted as a narrow nucleon resonance with the mass 1685 MeV, total width $\le 25$~MeV, and the photocoupling 
to the proton much smaller than to the neutron. Especially, the latter property is easily understood assuming that $N(1685)$ is
a cryptoexotic member of~$\overline{\boldsymbol{10}}$~\cite{Polyakov:2003dx}.

The argument for small proton coupling is based on the approximate \mbox{$U$-spin} sub-symmetry of flavor SU(3). Both $\eta$ and photon are 
$U$-spin singlets and neutron and proton are $U$-spin triplet and doublet, respectively. The neutron- and proton-like members of
$\overline{\bf 10}$ are $U$-spin triplet and 3/2 multiplet, respectively. Therefore, in the SU(3) symmetry limit, proton photo-excitation
to $p_{\overline{\bf 10}}+\eta$ is forbidden, while neutron transition to $n_{\overline{\bf 10}}+\eta$ is allowed. For alternative explanations,
see Refs.~\cite{Shklyar:2006xw,Anisovich:2008wd,Doring:2009qr}.

It was found that the width of $N(1685)$
is in the range of tens of MeV with a very small $\pi N$ partial width of
${\mit\Gamma}_{\pi N}\leq 0.5$~MeV~\cite{str}. One should stress that the
decay to $\pi N$ is not suppressed in the SU$(3)$ limit and it
can be made small only if the symmetry violation is taken into account. Therefore, in Ref.~\cite{Goeke:2009ae},
masses and widths of exotic $\overline{\bf 10}$ were reanalyzed taking into account the mixing of the ground-state
octet with antidecuplet, already discussed in Section~\ref{ssec:Thetawidth}, 
and antidecuplet mixing with the excited Roper resonance octet. Taking into account
all available data on different branching ratios and some model input, it was possible to constrain the mixing
angles\footnote{Due to the accidental equality of the SU(3) Clebsch--Gordan coefficients, the mixing angles of
${\mit\Sigma}$ and $N$ states in octet and decuplet are equal, so only two mixing angles were necessary for
the discussed mixing pattern.} leading to 
\begin{align}
1795\;\text{MeV}  &  <M_{{\mit\Sigma}_{\overline{10}}}<1830\;\text{MeV} \, ,%
\notag\\
1900\;\text{MeV}  &  <M_{{\mit\Xi}_{\overline{10}}}<1970\;\text{MeV} \, 
\label{xirange}%
\end{align}
with the decay widths
\begin{align}
9.7\;\text{MeV}  &  <{\mit\Gamma}_{{\mit\Sigma}_{\overline{10}}}<26.9\;\text{MeV} \, ,%
\notag\\
7.7\;\text{MeV}  &  <{\mit\Gamma}_{{\mit\Xi}_{\overline{10}}}<11.7\;\text{MeV} \, .
\label{xiGamma}%
\end{align}
These limits follow from the assumptions that ${\mit\Theta}^+$ mass is 1540~MeV and its width is 1~MeV, and 
that the decay width of $N(1685)$ is smaller than 25~MeV. One sees that the decay width of ${\mit\Xi}_{\overline{10}}$
is still small, but larger than in the SU(3) limit. Its mass is still in the range scanned by NA61/SHINE.

\vspace{-2mm}
\section{Experiments}
\label{sec:exps}

The positive evidence for ${\mit\Theta}^+$ by LEPS~\cite{LEPS:2003wug} and DIANA~\cite{DIANA:2003uet} 
has prompted a number of searches by other experimental groups. 
At that time, only data collected originally for searches other than ${\mit\Theta}^+$ was available. Only later were dedicated experiments designed 
and conducted. For a complete list of experiments, we refer the Reader to reviews from 2008~\cite{Danilov:2008uxa},  from 2014~\cite{Liu:2014yva}, to a 
general review of the strange baryon spectrum~\cite{Hyodo:2020czb} and to a recent paper~\cite{Amaryan:2022iij}. 

Below, we will briefly recall only a few experiments, mainly those that 
have so far upheld their initial positive results.

\vspace{-1.5mm}
\subsection{LEPS  and photproduction experiments}

Acronym LEPS stands for the Laser-Electron Photon facility at \mbox{SPring-8}, which is an electron storage ring located approximately
10 km NW from Himeji in Japan. The LEPS detector was optimized for measuring $\phi$-mesons produced near the threshold 
energy from a photo-production on a hydrogen target by detecting the $K^+ K^-$ pairs from the $\phi$ decays. Interestingly,
photons have been obtained from a Compton backscattering of electrons by a laser beam. The photon energy used in 
the analysis was $1.5~{\rm GeV}<E<2.4~{\rm GeV}$.

\begin{figure}[htb]
\centerline{%
\includegraphics[width=9.2cm]{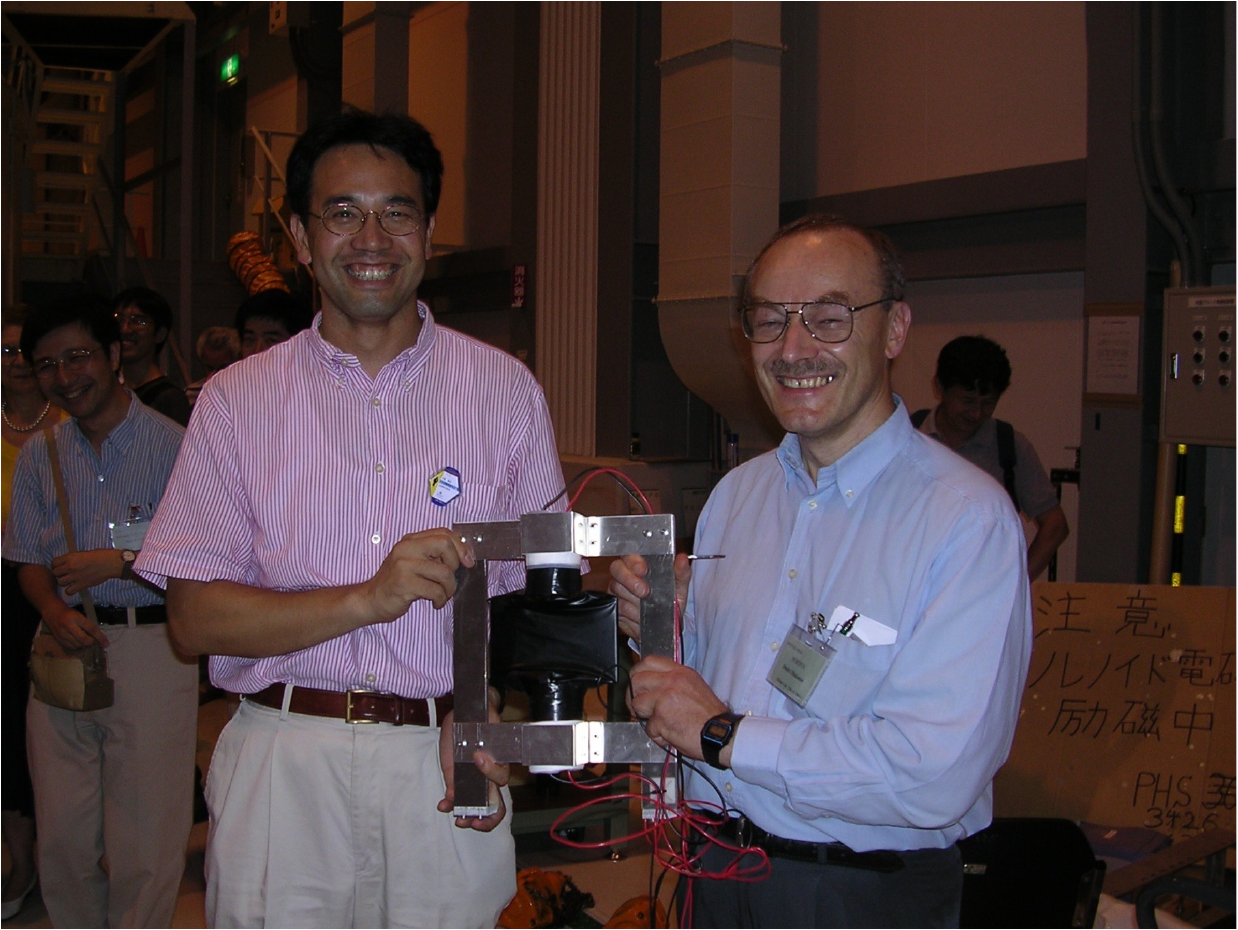}}
\caption{Takashi Nakano and Dmitry Diakonov holding a 0.5~cm thick plastic scintillator
that was used as a target in the LEPS experiment. (Photo taken at the Pentaquark Workshop at Spring-8 facility in 2004,
courtesy to the unknown author.)}%
\label{fig:TNMDsilicon}%
\end{figure}

\begin{figure}[htb]
\centerline{%
\includegraphics[width=5cm]{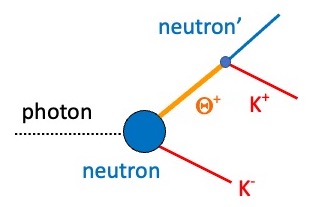}}\vspace{-1mm}
\caption{${\mit\Theta}^+$ photoproduction at LEPS.}%
\label{fig:LEPS}%
\end{figure}

Unfortunately, the hydrogen (proton) target was not an option for  ${\mit\Theta}^+$
production\footnote{Called $Z^+$ at the time.}
with the $K^+ K^-$ final state. 
Luckily, 9.5~cm behind the
liquid hydrogen target there was a so-called START counter (see Fig.~\ref{fig:TNMDsilicon}), a 0.5~cm thick plastic scintillator, 
which was composed of hydrogen and carbon nuclei, 
C:H~$ \approx 1:1$. The production of ${\mit\Theta}^+$ took place on a neutron inside of a carbon nucleus. The neutron
($n'$) from the pentaquark decay was not measured, Fig.~\ref{fig:LEPS}, so one looked for a signal in  the missing mass $M_{\gamma  K^-}=E_{\gamma}-E_{K^-}$
distribution, and --- for comparison --- in $M_{\gamma  K^+}=E_{\gamma}-E_{K^+}$. No pentaquark signal was 
detected in the latter case.

\newpage

The main problem was, however, that the target neutron was inside a carbon nucleus, and its momentum was smeared
by the Fermi motion. After applying the Fermi motion correction, the ${\mit\Theta}^+$ peak was clearly visible at $M_{{\mit\Theta}^+}=1.54\pm0.01$~GeV 
with 4.6$\sigma$ Gaussian significance. The width was estimated to be smaller than 25~MeV.

\vspace{-2mm}

Five years later, in 2008, LEPS published results from a dedicated photo-production experiment, this time on a deuteron target~\cite{LEPS:2008ghm}.
Although the measurement strategy was basically the same as in the case of carbon, the deuteron setup offered a possibility to cross-check the 
pentaquark production in a $\gamma n \rightarrow K^- {\mit\Theta}^+$ reaction with ${\mit\Lambda}(1520)$ production in $\gamma p \rightarrow K^+ {\mit\Lambda}^0(1520)$.
This was possible because the LEPS detector has a symmetric acceptance for positive and negative particles. In the analysis, the LEPS Collaboration
paid special attention to the uncertainties  related to the compositeness of the deuteron: the Fermi motion and the role of the spectator nucleon. To this end,
they developed a so-called minimum momentum spectator approximation. The analysis confirmed the existence of a narrow ${\mit\Theta}^+$ signal
at $M_{{\mit\Theta}^+}=1.524\pm0.002 \pm 0.003$~GeV. The significance was estimated to be 5.1$\sigma$ 
and the width was much smaller than $30$~MeV.

One should note that the  analysis was performed using the data collected with the LEPS detector in 2002--2003, where the statistics was  
improved by a factor of 8 over the previous measurement~\cite{LEPS:2003wug}. It took, however, five years to finally publish the results. The 
reason was probably that
in the meantime, a number of experiments reported negative results, and skepticism about the existence of ${\mit\Theta}^+$ was growing. Most importantly,
in the analogous experiment carried out by the CLAS (CEBAF
Large Acceptance Spectrometer\footnote{CEBAF stands for Continuous Electron Beam Accelerator Facility at Jefferson Laboratory
located in Newport News, VA, USA.}) 
Collaboration,
no narrow peak corresponding to ${\mit\Theta}^+$ was observed~\cite{CLAS:2006czw}, contradicting the earlier CLAS report from 2003~\cite{CLAS:2003yuj}.

The CLAS $\gamma d$ experiment was analogous to LEPS, but not identical. 
CLAS observed all charged particles in the final state, including the spectator
proton. This required an elastic  rescattering of $K^-$ (see Fig.~\ref{fig:LEPS}) off the proton (not shown in Fig.~\ref{fig:LEPS}),
so that the proton would gain enough momentum
to allow detection. The probability of such a rescattering was an essential factor in the CLAS analysis. Since LEPS assumed the proton to be a spectator,
the kinematic conditions of the two experiments were different. Moreover, the angular coverage of both detectors was also different:
less than 20 degrees for  LEPS and greater than 20 degrees for  CLAS in the LAB
system~\cite{Nakano:2017zrr}. 

To clarify the situation, the LEPS Collaboration performed the search for ${\mit\Theta}^+$  in the $\gamma d \rightarrow K^+ K^- n p$ reaction
with 2.6 times higher statistics. The peak was still there.
In 2013--2014, a new measurement was performed with the improved proton acceptance. Partial results were published
in different conference proceedings~\cite{Nakano:2017zrr,Nakano:2017fui,Yosoi:2019mno} but to the best of our knowledge, a full-fledged journal article
has not yet been released. 

At the end of 2022, the LEPS2 detector started to collect new data in the search for ${\mit\Theta}^+$~\cite{Nakanoprivate}. The LEPS2 detector has
better angular coverage than LEPS and will look for ${\mit\Theta}^+$ in the following  reactions~\cite{Ahn:2023hiu}: (1) 
$\gamma n \rightarrow K^- {\mit\Theta}^+$ and (2) $\gamma p \rightarrow {\bar{K}}^{0\,*} {\mit\Theta}^+$, where ${\mit\Theta}^+$ will be reconstructed from
the following decays: ${\mit\Theta}^+ \rightarrow p K^{0}_{\rm S}\rightarrow p\, \pi^+\pi^-$ and in the second case additionally 
${\bar{K}}^{0\,*}\rightarrow K^- \pi^+$. Apparently all four or five particles in the final state will be identified, which means that the uncertainty
of the previous measurements due to the Fermi motion of the target  neutron or proton will be removed. 
We therefore look forward to future results.

To circumvent the problem of small $g_{{\mit\Theta} NK}$~(\ref{eq56}), 
Amaryan with Diakonov and Polyakov~\cite{Amarian:2006xt} proposed in 2006 to look for ${\mit\Theta}^+$
at CLAS in the interference with the $\phi$ meson. The interference cross-section is linear in $g_{{\mit\Theta} NK}$
and therefore is larger than the production cross-section where the $\phi$ contribution is removed. The corresponding analysis
was published six years later~\cite{Amaryan:2011qc} with a positive result. Nevertheless, this paper has not been 
formally approved by the
entire CLAS Collaboration, which criticized kinematical cuts applied in~\cite{Amaryan:2011qc} 
and published an official disclaimer~\cite{CLAS:2012gcw}.

It is important to realize that the theoretical estimation of  photoproduction is hampered by uncertainties that concern a  
$\gamma n \rightarrow {\mit\Theta}^+ K^-$
vertex that in Fig.~\ref{fig:LEPS} is depicted as a large (blue) blob. The leading contribution corresponds to photon dissociation into two kaons
 $\gamma \rightarrow K^+ K^-$ followed
by a formation of a resonance $K^+ n \rightarrow {\mit\Theta}^+$.  The latter coupling can be estimated from the ${\mit\Theta}^+$ decay width, however, the photon
dissociation is less known. Moreover, a process involving $K^*$ is also possible:  $\gamma \rightarrow K^{+*} K^-$ 
followed by  $K^{+*} n \rightarrow {\mit\Theta}^+$.
Arguments have been brought up that $K^*$ contribution should be small, it however adds to an overall uncertainty. In the case of the photoproduction
on the proton, the same arguments apply to $\gamma \rightarrow {\bar{K}}^{0\,*}K^0$.

\vspace{-2.5mm}
\section{DIANA --- resonance formation}

\vspace{-.5mm}
Unlike photoproduction, resonance formation is the cleanest experiment  possible in the search for ${\mit\Theta}^+$. The Breit--Wigner
cross-section for the production of a resonance of spin $J$ and mass $M$ in the scattering of two hadrons of spin $s_1$ and $s_2$
takes the following form (see \eg\ Eq.~(51.1) in Ref.~\cite{ParticleDataGroup:2024cfk}):
\begin{equation}
\sigma_{\rm BW}(E)=\frac{2J+1}{(2s_1+1)(2s_2+1)}\frac{\pi}{k^2}B_{\rm in} B_{\rm out}\frac{{\mit\Gamma}^2}{(E-M)^2+{\mit\Gamma}^2/4}\,,
\end{equation}
where $E$ is the c.m. energy, $k$ is the c.m.\ momentum of the initial state, and ${\mit\Gamma}$ is the full width
at the half maximum height of the resonance. The branching fraction for the resonance into the initial-state channel is $B_{\rm in}$
and into the final-state channel is $B_{\rm out}$ --- in the present case for $K^+n$ scattering and one of the possible final
states $K^+n$ or $K^0 p$, we have   $B_{\rm in} =B_{\rm out}=1/2$. Substituting the ${\mit\Theta}^+$ mass, one gets
that the cross-section at the peak $\sigma_{\rm BW}(M_{{\mit\Theta}^+})\sim 15\div 20$~mb. This is a model-independent prediction,
and we see that the  cross-section for the ${\mit\Theta}^+$ production in $KN$ scattering is large. A more detailed study of 
the ${\mit\Theta}^+$ production in the $K^+d \rightarrow K^0pp$ reaction shows that the production cross-section is in this
case of the order of 5~mb~\cite{Sekihara:2019cot}. The pertinent feasibility study of searching for ${\mit\Theta}^+$ in this channel at J-PARC was
recently performed in Ref.~\cite{Ahn:2023hiu}.

The formation process was used in the DIANA experiment where the bubble chamber DIANA filled with liquid xenon 
was exposed to a  $K^+$ beam  from the ITEP proton synchrotron. In Ref.~\cite{DIANA:2003uet}, the authors
analyzed the $K^0 p$ effective mass spectrum in the $K^+n \rightarrow K^0 p$ reaction on a nucleon bound in a xenon
nucleus. A resonant enhancement with $M = 1539 \pm 2$~MeV$/c^2$ and ${\mit\Gamma} \le 9$~MeV$/c^2$ was
observed. The statistical significance of the enhancement  was estimated to be 4.4$\sigma$.

The DIANA Collaboration continued analysis of the bubble chamber films and in 2006 published new results from the
larger statistics sample~\cite{DIANA:2006ypd}. They confirmed their initial observation with the mass of $M = 1537 \pm 2$~MeV$/c^2$ with,
however,
a much smaller estimate of the width: ${\mit\Gamma} = 0.36\pm 0.11$~MeV$/c^2$. Depending
on the significance estimator, they obtained the statistical significance of 4.3, 5.3 or 7.3$\sigma$.
Three years later they increased again statistics confirming the existence of ${\mit\Theta}^+$ with approximately
the same mass and width, but higher statistical significance reaching 8$\sigma$~\cite{DIANA:2009rzq}. These results
were confirmed in their last publication from 2014~\cite{DIANA:2013mhv}.

In 2006, the Belle Collaboration reported search results for ${\mit\Theta}^+$~\cite{Belle:2005thz}.
In the Belle experiment located at the KEKB asymmetric  collider, interactions of secondary particles with detector 
material were used to search for~${\mit\Theta}^+$. Belle performed two different analyses.
In the first one, they searched for inclusive production of ${\mit\Theta}^+$ in the
$K N \!\rightarrow\! {\mit\Theta}^+ X$ reaction with a subsequent decay ${\mit\Theta}^+ \!\rightarrow\! p K^0_{\rm S}$,
using the signal from
inclusive  ${\mit\Lambda} (1520)$ production as a reference. In the second one, they looked at exclusive ${\mit\Theta}^+$
production in the charge exchange reaction  $K^+n \!\rightarrow\! {\mit\Theta}^+ \!\rightarrow\! p K^0_{\rm S}$. The latter one
is directly comparable with the DIANA experiment.
No formation signal of the ${\mit\Theta}^+$ baryon was observed, and an upper limit on the ${\mit\Theta}^+$  width was estimated:
${\mit\Gamma} < 0.64$~MeV for $M_{{\mit\Theta}^+} = 1539$~MeV.
 
One of the reasons for the skepticism about ${\mit\Theta}^+$ were the above results for its unnaturally --- as it seemed at the time --- small width.
In soliton models, as explained in Section~\ref{ssec:Thetawidth}, there are natural mechanisms that lead to very small pentaquark widths.
Here, let us only mention that recently found by the LHCb Collaboration at CERN excited ${\mit\Omega}_c(3050)$ has a total
width ${\mit\Gamma}=0.8\pm 0.2 \pm 0.1$~MeV$/c^2$~\cite{Aaij:2017nav}.  
In a later publication from 2021~\cite{LHCb:2021ptx} the LHCb Collaboration concluded that: {\em  The natural width of the
${\mit\Omega}_c^0(3050)$ is consistent with zero}.
${\mit\Omega}_c(3050)$ was found in the decay to ${\mit\Xi}_c K^-$~\cite{Aaij:2017nav}, where
the kaon momentum is $p=275$~MeV$/c^2$. This is approximately $\sim 10$~MeV$/c^2$ above the kaon momentum in the decay
of ${\mit\Theta}^+$. From this perspective, the small pentaquark width is not particularly ``unnatural''. As a consequence, the small width of ${\mit\Omega}_c(3050)$
led to its interpretation as a heavy charm pentaquark belonging to the exotic SU(3) $\overline{\bf 15}$ 
multiplet~\cite{Kim:2017khv,Kim:2017jpx,Praszalowicz:2022hcp}, see Section~\ref{ssec:HBs}.

The formation experiment with the $K^+$ beam can be easily performed at the J-PARC facility in Japan
looking at the three-body final-state $K^0 p p$~\cite{Ahn:2023hiu}.
Another very promising search for ${\mit\Theta}^+$ will be possible at the already approved program at the $K_\mathrm{L}$ facility  at JLab
\cite{Amaryan:2022iij,KLF:2020gai,Amaryan:2024koq}. Here, with a secondary beam of kaons, one may look at
a two-body reaction $K^0_\mathrm{L} p \rightarrow K^+ n $ on the hydrogen target. The plan is to measure
the initial energy  benefiting from the design momentum resolution below 1 MeV rather
than the invariant mass of the $K^+n$ system. 
According to the current schedule, data collection will start in 2026~\cite{Amaryan:2024koq}.
Note that the two-body final state is much cleaner than the three-body one,
which is proposed to be studied at J-PARC. Finally, at the $K_\mathrm{L}$ facility, one will  also 
be able to look for other members of antidecuplet, like ${\mit\Xi}^+$.

If ${\mit\Theta}^+$ exists, it should be visible in partial wave analyses (PWA) of $K^+N$ scattering. In Ref.~\cite{Strakovsky:2024ppo}
in this volume, you may find a description of the modifications needed to see a very narrow structure in the PWA
and the results.

\section{Summary}

The story of ${\mit\Theta}^+$ is not only interesting for physics. It is like a detective story with unexpected twists, 
where we do not know if the victim is alive or dead, or even if it existed at all.
It is a story about enthusiasm for an epochal discovery, a story about fast and optimistic shortcuts, and 
a painful return to reality. It is a story of emotions --- positive and negative. While it seems that ${\mit\Theta}^+$
and light baryonic exotica research is presently on hold, we should expect some new experimental results
in not-so-distant future.

\vspace{7mm}
I owe a lot to Mitya, Vitya, and Maxim with whom I explored possibilities for exotica. I am indebted
to Hyun-Chul Kim for a longstanding collaboration, and to Igor Strakovsky and Moskov Amaryan for keeping me
informed about experimental searches for ${\mit\Theta}^+$.

\flushleft

\end{document}